\documentclass[journal=jctcce,manuscript=article]{achemso}
\setkeys{acs}{usetitle=true}

\usepackage[version=3]{mhchem} 

\usepackage[normalem]{ulem}

\usepackage{color, xcolor, colortbl}
\usepackage{graphicx,epstopdf}
\usepackage{geometry}
\usepackage{amsmath,amssymb,amsthm}
\usepackage[ruled,vlined]{algorithm2e}
\usepackage{bm}
\usepackage[caption=false]{subfig}
\usepackage{appendix}
\usepackage{multirow}
\usepackage{braket}
\usepackage[english]{babel}
\usepackage{hyperref}
\usepackage[capitalize]{cleveref}
\usepackage{adjustbox}
\usepackage{xspace}
\usepackage[roman]{complexity}
\usepackage{comment}

\newcommand{\diag}{\operatorname{diag}}

\newcommand{\mc}[1]{\mathcal{#1}}

\newcommand{\LL}[1]{\textcolor{brown}{[LL:#1]}}
\newcommand{\FF}[1]{\textcolor{blue}{[FF:#1]}}

\newcommand{\ed}[1]{\textcolor{black}{#1}}

\title{On the pure state $v$-representability of density matrix embedding theory}

\author{Fabian M. Faulstich}
\affiliation{Department of Mathematics, University of
California, Berkeley, California 94720, United States}
\altaffiliation{These authors contributed equally.}

\author{Raehyun Kim}
\affiliation{Department of Mathematics, University of
California, Berkeley, California 94720, United States}
\altaffiliation{These authors contributed equally.}

\author{Zhi-Hao Cui}
\affiliation{Division of Chemistry and Chemical Engineering, California Institute of Technology, Pasadena, California 91125, United States}

\author{Zaiwen Wen}
\affiliation{Beijing International Center for Mathematical Research, BICMR, Peking University, Beijing, China}

\author{Garnet Kin-Lic Chan}
\affiliation{Division of Chemistry and Chemical Engineering, California Institute of Technology, Pasadena, California 91125, United States}

\author{Lin Lin}
\email{linlin@math.berkeley.edu}
\affiliation{Department of Mathematics, University of
California, Berkeley, California 94720, United States}
\altaffiliation{Computational Research Division, Lawrence Berkeley National Laboratory,
Berkeley, California 94720, United States}

\begin{document}

\begin{abstract}

Density matrix embedding theory (DMET) formally requires the matching of density matrix blocks obtained from high-level and low-level theories, 
but this is sometimes not achievable in practical calculations. 
In such a case, the global band gap of the low-level theory vanishes, and this can require additional numerical considerations. 
We find that both the violation of the exact matching condition and the vanishing low-level gap are related to the assumption that the high-level density matrix blocks are non-interacting pure-state $v$-representable (NI-PS-V), which assumes that  the low-level density matrix is constructed following the Aufbau principle. 
In order to relax the NI-PS-V condition, we develop an augmented Lagrangian method to match the density matrix blocks without referring to the Aufbau principle. 
Numerical results for 2D Hubbard and hydrogen model systems indicate that in some challenging scenarios, the relaxation of the Aufbau principle directly leads to exact matching of the density matrix blocks, which also yields improved accuracy.
\end{abstract}

\section{Introduction}\label{sec:intro}


Density matrix embedding theory (DMET)~\cite{DMET2012, DMET2013, tsuchimochi2015density, bulik2014density, wouters2016practical, cui2019efficient,sun2020finite,cui2020ground} is a quantum embedding theory designed to treat strong correlation effects in large quantum systems. 
DMET and its related variants have been successfully applied to a wide range of systems such as Hubbard  models~\cite{DMET2012, bulik2014density, chen2014intermediate, boxiao2016, Zheng2017, zheng2017stripe, welborn2016bootstrap,senjean2018site,senjean2019projected}, quantum spin models~\cite{Fan15, gunst2017block,ricke2017performance}, and a number of strongly correlated molecular and periodic systems~\cite{DMET2013, wouters2016practical,cui2020ground,nusspickel2021systematic,bulik2014electron,pham2018can,hermes2019multiconfigurational,tran2019using,ye2018incremental,ye2019bootstrap,ye2019atom,ye2020bootstrap,ye2021accurate,tran2020bootstrap}. 
The main idea of DMET is to partition the global quantum system into several ``quantum impurities''. 
Each impurity is treated accurately via a high-level theory (such as full configuration interaction (FCI)~\cite{knowles1984new,olsen1990passing,vogiatzis2017pushing}, coupled cluster theory~\cite{Cizek1966}, density matrix renormalization group (DMRG)~\cite{White92}, etc.).
Global information, in particular the one-electron reduced density matrix (1-RDM), is made consistent between  all the impurities with the help of a low-level Hartree--Fock (HF) type of theory. 
In the self-consistent-field DMET (SCF-DMET)\footnote{Throughout the paper, DMET refers to SCF-DMET. This is in contrast to one-shot DMET, in which the impurity problem is only solved once without self-consistent updates.}, this global information is then used to update the impurity problems in the next self-consistent iteration, until a certain consistency condition of the 1-RDM is satisfied between the high-level and low-level theories~\cite{DMET2012,DMET2013,tsuchimochi2015density,bulik2014density,wu2019projected,wu2020enhancing}.

In DMET, the self-consistency condition can be achieved by optimizing a correlation potential, which can be viewed as a Lagrange multiplier associated with the matching condition of the 1-RDMs. 
For instance, if the self-consistency condition only requires electron densities from the high-level and low-level theories to match (e.g. in Ref.~\citenum{bulik2014density}), then the problem of finding the correlation potential \ed{strongly resembles} the $v$-representability problem in density functional theory (DFT)~\cite{HohenbergKohn1964,Levy1979,EnglischEnglisch1983,PerdewLevy1985,vanLeeuwen2003}.
Omitting the spin degree of freedom, an electron density $\rho$ (often obtained from a many-body calculation) with $N$ electrons is called non-interacting pure-state $v$-representable (NI-PS-V), if $\rho$ can be reconstructed (1) from a single particle Hamiltonian with potential $v$ (2) using the energetically lowest $N$ orbitals.
The condition~(2) is also referred to as the Aufbau principle. 
There are densities that are not NI-PS-V, but for DFT such densities are rare exceptions rather than the norm~\cite{vanLeeuwen2003}.

DMET requires the matching condition for certain 1-RDM matrix blocks corresponding to the high-level 1-RDMs.
Then the correlation potential (denoted by $u$ following the convention in the literature) consists of matrix blocks of matching dimensions. \ed{While $v$-representability in DFT usually concerns a diagonal potential in the real-space basis, the correlation potential in DMET is expressed as a block diagonal matrix in the fragment-orbital basis. }
In a typical DMET calculation, the 1-RDM is assumed to be NI-PS-V, in particular, the low-level 1-RDM is reconstructed following the Aufbau principle. 
However, from the very beginning of the development of DMET, it was noticed that the exact matching of the 1-RDMs often cannot be achieved~\cite{DMET2012,DMET2013,bulik2014density}.
Therefore, as a practical solution, the matching condition is relaxed 
into a least-squares procedure, i.e. finding a correlation potential that produces a low-level 1-RDM that is as close as possible to the high-level 1-RDM. 
Such a least-squares procedure is a non-convex optimization problem, and suffers from two robustness issues: (1) The objective function can have multiple local minima, and the optimization procedure may not converge to the global minimum. (2) The low-level problem can become gapless (i.e. there is no gap between the $N$-th and $(N+1)$-th eigenvalues), and the Aufbau principle becomes ill-defined.
Problem (1) can be solved by reformulating the correlation potential fitting as a semidefinite programming (SDP) problem~\cite{wu2020enhancing}, which is a convex optimization and is typically a robust procedure.
The gapless problem (2), however, is an intrinsic problem in DMET.
Proposition 2 in Ref.~\citenum{wu2020enhancing} shows that under suitable conditions, whenever the exact matching condition cannot be satisfied (i.e. the high-level 1-RDM blocks are not NI-PS-V), the low-level gap must also vanish.
 
In this paper, we present three examples that violate the NI-PS-V assumption, and which yield a gapless low-level Hamiltonian.
To overcome the numerical difficulties connected to the vanishing low-level gap, we suggest a modification of DMET that relaxes the NI-PS-V assumption: for a correlation potential $u$, the low-level 1-RDM (normalized and idempotent) is constructed to minimize the energy of the low-level Hamiltonian subject to exact matching, but using orbitals following \textit{any} occupation profile.
The idempotency condition implies that each occupation number can still only be $0$ or $1$, but the occupation profile may or may not follow the Aufbau principle.
This may seem a daunting problem, as the possible number of distinct occupation profiles is combinatorially large. 
We propose to use an augmented Lagrangian method~\cite{Hestenes1969,Powell1972method,nocedal2006numerical} (ALM), coupled with a projected gradient descent method, to solve this modified constrained optimization problem, which yields both the low-level 1-RDM and the correlation potential. 
We also propose a method to efficiently determine the occupation profile of the converged solution, which allows us to visualize the degree to which the Aufbau principle is violated. Numerical results indicate that this procedure exactly satisfies the matching condition even in cases when an exact fit following the Aufbau principle is not possible.

The rest of the paper is organized as follows. We first provide a brief review of DMET including the optimization problem describing the fit of the high-level 1-RDM by means of a low-level 1-RDM in Sec.~\ref{sec:DMETIntro}. In Sec.~\ref{sec:alm} we then introduce the augmented Lagrangian method to match the 1-RDM blocks without the Aufbau constraint. In Sec.~\ref{sec:numer} we present numerical evidence that the NI-PS-V condition of the high-level 1-RDMs in DMET can be violated, and report the performance of the augmented Lagrangian method for such cases in the 2D Hubbard model (Sec.~\ref{sec:2d_hub}), a linear hydrogen chain with large fragments (Sec.~\ref{sec:H36}), and an H$_6$ model at a range of geometries (Sec.~\ref{sec:H6}).

\section{Theory}\label{sec:theory}

\subsection{Brief review of density matrix embedding theory}
\label{sec:DMETIntro}

Consider the electronic structure Hamiltonian in second quantization, i.e. 
\begin{equation}
\label{eq:secQuanHam}
H = \sum_{pq}^L t_{pq} a_p^\dagger a_q + \frac{1}{2}\sum_{pqrs}^L v_{prqs} a_p^\dagger a_q^\dagger a_s a_r,
\end{equation}
where $t_{pq}$ and $v_{prqs}$ characterize the one- and two-particle interactions, respectively, and $L$ is the number of orbitals---subsequently spin is omitted, but the discussion directly generalizes to spin restricted/unrestricted settings. We will refer to problems defined over the full set of $L$ orbitals as global problems.
The goal is to determine expectation values of the global eigenstate $\Psi$ of $H$ (here assumed the ground-state) when  $L$ is too large for a high-level determination of $\Psi$ to be practical.
The idea of DMET is to reduce the problem size by reformulating it as a collection of quantum impurity (fragment) problems (labelled by $x$), where the impurity problems individually contain a small number of orbitals, such that their ground-state can be determined to high-accuracy. The wavefunctions of the impurity problems, $\Psi_x$, will be called the high-level solutions. DMET defines both a procedure to construct the impurity problems, as well as how to assemble the information from the individual $\Psi_x$ to approximate expectation values of $\Psi$.  

In SCF-DMET, the impurities contain $A_x$ orbitals (chosen as non-overlapping subsets of the $L$ orbitals), and each impurity is augmented by a set of $B_x$ bath orbitals. The bath orbitals are obtained via a self-consistent procedure that is based on an approximate solution of the 
global problem, the ground-state $\Phi$ of an auxiliary low-level global Hamiltonian
\begin{equation}
\label{eq:LL-Ham}
H^{\rm ll}(u)  = f + c(u),
\end{equation}
where $f$ is a mean-field Fock matrix
and $c(u) = \sum_{p,q}u_{p,q}a_p^\dagger a_q$ is an effective single-body interaction known as the correlation potential $u$. Because of the mean-field form of $\Phi$, it can be characterized entirely by its 1-RDM $D$, and the bath orbitals are defined via a singular value decomposition of a sub-block of $D$ (for a complete discussion see Refs.~\citenum{wouters2016practical,wu2020enhancing}, and for additional work, including some that goes beyond assuming a mean-field $\Phi$, see Refs.~\citenum{fertitta2018rigorous,nusspickel2020frequency,nusspickel2020efficient,sriluckshmy2021fully,tsuchimochi2015density,fertitta2019energy,ye2019atom,ye2019bootstrap,ye2020bootstrap,ye2018incremental,tran2020bootstrap}). The correlation potential $u$ is determined by a self-consistent matching procedure and this is the mathematical problem we seek to address.

Specifically, here we investigate the impurity density matrix matching problem ~\cite{wouters2016practical}, where the global high-level 1-RDM (assembled from the high-level solutions of the impurity problems by democratic partitioning) is matched with
the low-level matrix blocks that correspond to the individual fragments. 
More precisely, the matching condition is $D_x = P_x$ where $D_x$ and $P_x$ are the matrix blocks corresponding to the individual fragments in the low-level and high-level 1-RDM, respectively.
In the following, we assume w.l.o.g. that each fragment has the same size $A_x$ 
and that the fragment orbitals are numbered consecutively.
The matching condition then yields the following constrained optimization problem
\begin{equation}
\label{Eq:OptProb}
\left\lbrace
\begin{split}
&\min_{D\in\mathbb{R}^{L\times L}} &&{\rm Tr} (f D),\\
&\quad {\rm s.t. } &&D \in \mathcal{M} ~{\rm and}~ D_x = P_x~~  \forall x,
\end{split}
\right.
\end{equation}
and $\mathcal{M}$ denotes the set of admissible density matrices, i.e. 
\begin{equation}
\mathcal{M} = \{ D \in\mathbb{R}^{L\times L} ~|~ D =D^T,\, {\rm Tr}(D) = N,\, D^2 = D\}.
\end{equation}

The standard optimization algorithm in SCF-DMET writes the low-level 1-RDM as the ground-state 1-RDM of $H^\mathrm{ll}(u)$, and then formulates the optimization problem in Eq.~\eqref{Eq:OptProb} as a least squares minimization of $\sum_x ||D_x(u)- P_x||_F^2$ (cf. Eq.~(8) in Ref.~\citenum{wu2020enhancing}).
This corresponds to an additional constraint on the approximate density matrix $D$, namely
\begin{equation}
\label{eq:Aufbau}
D = C C^\dagger
\end{equation}
where $C\in\mathbb{C}^{L \times N}$ is the orbital coefficient matrix of the ground state Slater determinant of $H^{\mathrm{ll}}(u)$.
If the cost function of the least-squares procedure is $0$ at the minimizer, the
underlying assumption 
is that there exists an auxiliary non-interacting system for which the ground-state 1-RDM (following the Aufbau principle) describes exactly the high-level 1-RDM blocks. As mentioned in the introduction, this is the non-interacting pure-state $v$-representability (NI-PS-V) condition in DFT~\cite{vanLeeuwen2003} \ed{except that the correlation potential is not necessarily diagonal in the real-space basis}. Typical DMET calculations assume that \ed{(this version of) NI-PS-V is satisfied for the high-level density blocks}, but the validity of this assumption has not been carefully scrutinized\footnote{Sometimes finite temperature smearing is used to generate the 1-RDM, then technically the 1-RDM is  {\it not} generated from a non-interacting pure state. This can sometimes improve the numerical convergence of DMET. When finite temperature smearing is used, the exact matching condition is often violated.}. 
\ed{In order to focus on the NI-PS-V assumption in DMET and provide clear numerical results, we moreover do not consider additional relaxation effects in the full system Fock matrix, i.e. the full charge self-consistency. The full charge self-consistency introduces an implicit change of the non-local potential from the updated density matrix, which is beyond the definition of NI-PS-V.}

\subsection{Fitting without obeying the Aufbau principle: augmented Lagrangian method }
\label{sec:alm}

In the context of SCF-DMET, the procedure for correlation potential fitting is a crucial 
step for numerical robustness
and different approaches have been proposed~\cite{wouters2016practical,wu2020enhancing,wu2019projected,ye2020bootstrap,ye2021accurate,nusspickel2021systematic,sriluckshmy2021fully,tsuchimochi2015density,fertitta2019energy}. 
One place where numerical issues can arise is when the global low-level Hamiltonian becomes gapless, in which case the least-squares cost function becomes non-differentiable. Recent analysis where the constrained optimization in Eq.~\eqref{Eq:OptProb} is reformulated as a convex optimization~\cite{wu2020enhancing} sheds light on this behaviour. Namely, under mild conditions, the convex optimization must yield a solution of Eq.~\eqref{Eq:OptProb} with exact matching \emph{unless} the low-level gap vanishes, in which case the set of constraints is inconsistent. From this, we conclude that the lack of exact matching, vanishing of the low-level gap, and violation of NI-PS-V all occur simultaneously---a connection which to the best of our knowledge has not previously been drawn. Although a particular algorithm, such as least-squares optimization, may still return a solution in this case, the non-zero error in the cost function is fundamentally unavoidable as a consequence of violating NI-PS-V.


We now suggest an approach that fits the high-level 1-RDM directly (as opposed to an indirect fit by means of optimizing the correlation potential) but which does not assume NI-PS-V of the high-level 1-RDMs. In other words, we relax the condition in Eq.~\eqref{eq:Aufbau} to allow for a low-level 1-RDM construction from orbitals following any occupation profile. We emphasize that the subsequent construction yields an energetic minimum, i.e. the fitted 1-RDM minimizes ${\rm Tr}(fD)$ over $\mathcal{M}$ while fulfilling the matching conditions. 

The presented approach to this optimization problem is based on the {\it augmented Lagrangian method}~\cite{nocedal2006numerical}. 
It was originally proposed to circumvent numerical difficulties that arise in the quadratic penalty method 
in the large penalty-parameter limit~\cite{Hestenes1969,powell1978algorithms}, and has since proven to be a very useful numerical tool for solving constrained optimization problems.
Recall that a constrained optimization problem can be replaced by an unconstrained optimization problem that includes an additional term penalizing the violation of the desired constraint. Intuitively, the simplest way to apply this idea to the optimization problem in Eq.~\eqref{Eq:OptProb} is to introduce a quadratic penalty term to the objective function, i.e. the original constrained optimization problem in Eq.~\eqref{Eq:OptProb} can be expressed as an unconstrained minimization problem of the function, e.g. 
\begin{equation}
\label{eq:QuadPen}
Q(D, \alpha)
=
{\rm Tr} (fD) +
\frac{\alpha}{2}
\sum_x 
\Vert D_x -P_x \Vert_F^2,
\end{equation}
where $\Vert \cdot \Vert_F$ is the Frobenius norm, and $\alpha$ is a penalty parameter.
By sequentially increasing $\alpha$, i.e. replacing $\alpha$ by an increasing sequence $(\alpha_k)$, the constraint violations become more severely penalized, and thereby force the minimizer of the penalty function to approach the feasible region for the constrained problem. Hence, in order to fulfill the constraint, $Q$ needs to be considered in the large parameter limit, i.e. $\alpha_k\to \infty$. The major drawback of this penalty approach is that the minimization of $Q(\cdot , \alpha_k)$ becomes in general more difficult as $\alpha_k$ becomes larger, as the Hessian can become ill-conditioned near the minimizer~\cite{nocedal2006numerical}. 
Aside from numerical poor performance, which may be overcome by careful case-by-case considerations~\cite{nocedal2006numerical}, it is straightforward to see that the quadratic penalty method does in general not fulfill a first order condition, i.e. at the exact solution $D_*$ of the original optimization problem in Eq.~\eqref{Eq:OptProb}, the gradient $\nabla_D Q(D_*,\alpha)$ is not zero~\cite{powell1978algorithms}. 

Alternatively, we may approach the constrained optimization problem in Eq.~\eqref{Eq:OptProb} with the method of Lagrange multipliers, i.e. finding the stationary points of
\begin{equation}
\label{eq:Lagrange}
\mathcal{L}(D,\{u_x\})=
{\rm Tr} (fD) +
\sum_x
{\rm Tr} (u_x(D_x - P_x)).
\end{equation}
Although the method of Lagrange multipliers fulfills the first order condition, the stationary points of $\mathcal{L}$ are always saddle points, which may complicate the numerical optimization procedure. 

We can overcome the challenges associated with the saddle point by adding a quadratic penalty term (an {\it augmentation}) making the objective function strongly convex~\cite{powell1978algorithms}. This is the augmented Lagrangian method going back to Hestenes~\cite{Hestenes1969}, Powell~\cite{Powell1972method}, and Rockafellar~\cite{rockafellar1973dual}. The augmented Lagrangian for the optimization problem in Eq.~\eqref{Eq:OptProb} then reads
\begin{equation}
\label{eq:Aug_lag}
\mathcal{L}_\alpha(D, \{u_x\})
=
{\rm Tr} (fD) +
\sum_x \left(
{\rm Tr} (u_x(D_x - P_x))+ \frac{\alpha}{2} \Vert D_x -P_x \Vert_F^2
\right),
\end{equation}
where $\{u_x\}$ corresponds to the Lagrange multipliers,
and $\alpha >0$ is a penalty parameter. Note that we may also refer to $u = {\rm diag} (\{u_x\})$ as the Lagrange multiplier, and the underlying connection to the correlation potential is elaborated in Sec.~\ref{sec:postprocessing}. At first glance, ALM simply mixes the Lagrange multiplier and quadratic penalty methods. However, an important advantage of ALM is that $\alpha$ does not have to be considered in the infinite parameter limit, instead, under mild conditions, a threshold parameter $\bar{\alpha}$ can be established, so that when $\alpha>\bar{\alpha}$, the minimizer of Eq.~\eqref{Eq:OptProb} (denoted by $D_*$) is a strict local minimizer of the augmented Lagrangian $\mc{L}_{\alpha}$ (cf. Theorem 17.5 in Ref.~\citenum{nocedal2006numerical}). Hence, a systematic increase of $\alpha_k$ towards some finite value, so that $\lim_{k \to \infty }\alpha_k > \bar{\alpha} $ is sufficient to fulfill the constraint which avoids the ill-conditioning of the Hessian in the infinite penalty parameter limit. 
\ed{
If the distance between the estimated and exact Lagrange multipliers is controlled suitably and the penalty parameter is bounded, then the convergence rate of the Lagrange multipliers is $Q$-linear.
Furthermore, any accumulation point generated by the algorithm converges to a stationary point under certain assumptions (for more details see Proposition 2.7 in~Ref.~\citenum{bertsekas2014constrained}). 
The upper bound of the step size $t$ (vide infra) is usually the reciprocal of the Lipschitz constant of the gradient of the Lagrangian function.}

\ed{Proposition 2 in Ref.~\citenum{wu2020enhancing} shows that under mild conditions there exists a  correlation potential $u^\star$ solving the convex optimization problem, and the corresponding 1-RDM $D^\star$ follows the Aufbau principle.
If the low-level Hamiltonian $h= f + u^{\star}$ is gapped, then exact matching of the diagonal blocks of the 1-RDM can be achieved. 
Furthermore, if $u^{\star}$ is unique\footnote{The uniqueness is up to a constant shift, which does not change the 1-RDM. The precise condition for which $u^{\star}$ is unique appears to be subtle and is currently an open question (see Ref.~\citenum{wu2020enhancing}).}, then we know that $u^\star=u_{\rm alm}={\rm diag}(\{u_x\})$ and there exists a $D_{\rm alm}$ such that ${\rm Tr }(f D_{\rm alm})\leq {\rm Tr }(f D^\star)$. Here we used the linearity of the trace and that $\mathcal{M}$ contains in particular the Aufbau solutions. On the other hand, since $D^\star$ follows the Aufbau principle we also know that ${\rm Tr }((f + u^{\star})D^\star)\leq {\rm Tr }((f + u^{\star}) D_{\rm alm})$. The exact matching condition implies ${\rm Tr }(u^{\star}D^\star)= {\rm Tr }(u^{\star} D_{\rm alm})$, and we have ${\rm Tr }(f D^\star)\leq {\rm Tr }(f D_{\rm alm})$.
Therefore, ${\rm Tr} (f D_{\rm alm})= {\rm Tr }(f D^\star)$ and  $D_{\rm alm} = D^\star$, i.e., in this case the minimizer of Eq~\eqref{eq:Aug_lag} is identical to the solution of the convex optimization problem in Ref.~\citenum{wu2020enhancing} and satisfies the Aufbau principle.} 



We now minimize the augmented Lagrangian in Eq.~\eqref{eq:Aug_lag} following the iteration rule 
\begin{subequations}
\begin{equation}
\label{Eq:SubProb1}
D^{(k+1)} = \underset{D \in \mathcal{M}}{\rm argmin}~ \mathcal{L}_{\alpha}(D, \{u_x^{(k)}\} ),
\end{equation}
\begin{equation}
u_x^{(k+1)} = u_x^{(k)}+ \alpha (D_x^{(k+1)} -P_x),
\end{equation}
\end{subequations}
where generally we may set $u_x^{(0)} = 0$ for all $x$.

We solve the subproblem in Eq.~\eqref{Eq:SubProb1} by using the projected gradient method~\cite{nocedal2006numerical}. To that end, we define the corresponding gradient descent  step starting from $D^{(k+1,0)}= D^{(k)}$, i.e. 
\begin{equation}
\label{ConjGradStep}
W^{(\ell)} = D^{(k+1,\ell)} - t\nabla_D \mathcal{L}_\alpha[D^{(k+1,\ell)},\{u_x^{(k)}\}]
\end{equation}
with step size $t$, and project onto the matrix manifold $\mathcal{M}$ to obtain the next iteration step, i.e.
\begin{equation}
\label{Eq:ProjGrad}
\begin{aligned}
D^{(k+1,\ell+1)}
&=
\underset{D \in \mathcal{M}}{\rm argmin} \Vert D-W^{(\ell)}\Vert_{F}^2.
\end{aligned}
\end{equation}
Let the eigenvalue decomposition of $W^{(\ell)}$ be given by $Q^\dag {\rm diag}(\{w^{(\ell)}_x\}) Q$, 
then, the optimal solution of Eq.~\eqref{Eq:ProjGrad} is given by 
\begin{equation}
\label{eq:projGradStep}
D^{(k+1,\ell+1)}
=
Q^\dag {\rm diag}(\{d^{(\ell)}_x\}) Q,
\end{equation}
where $\{d^{(\ell)}_x\}$ is the optimal solution of
\begin{equation}
\label{eqn:scalar_min}
\begin{aligned}
&\min_{d\in\mathbb{R}^L} &&\Vert d- w^{(\ell)} \Vert_2^2
&\quad {\rm s.t. } && \sum_x d_x = N ~{\rm and}~ d_x \in \{0,1\}~ \forall x.
\end{aligned}
\end{equation}
Note that the solution to Eq.~\eqref{eqn:scalar_min} can be constructed from $w^{(\ell)}$ by replacing its $N$ largest elements by one and the remaining elements by zero. \ed{This projection ensures that the approximate density matrix is idempotent.}
We refer to the incorporation of the above augmented Lagrangian optimization into DMET as alm-DMET.

\ed{For the sake of simplicity, we subsequently refer to the SCF iterations as {\it DMET iterations} (or simply {\it iterations}). The additional optimization iterations per DMET iteration, i.e. outer while-loop iterations in the ALM algorithm (see Alg.~\ref{alg:ALM2}), are referred to as {\it outer iterations}. 
In the case of alm-DMET we require to perform additional projected gradient iterations per outer iteration; we refer to these iterations as {\it inner iterations}.}
An important numerical component is the choice of the parameters $\alpha$ and $t$. In the subsequent numerical simulations, we set the step length $t = 0.001$. For the penalty parameter $\alpha$, we start with the initial guess $\alpha= 0.001$, and update every 100 \ed{outer} iterations according to $\alpha_{k+1} = \frac{3}{2}\alpha_k$. The maximal value for the penalty parameter is set to $\alpha_{\rm max} = 10$. The numerical performance suggests that $\alpha_{\rm max} \geq \bar{\alpha}$ for the systems considered here. With said hyperparameter settings, we typically observe convergence within 20000 \ed{outer} iterations of the ALM algorithm. 
If $t$ is chosen to be too large, the algorithm may fail to converge. \ed{We emphasize that the number of outer iterations in the case of alm-DMET can have a sensitive dependence on the choice of hyperparameters. The choice of hyperparameters outlined above, which is used in the subsequent experiments, are chosen such that they can be applied to {\it all} systems presented in Section~\ref{sec:numer}, but are by no means optimal with respect to the number of outer iterations (see Appendix~\ref{app:mactroIts}). 
}

\ed{We highlight the importance of the choice of the initial density matrix $D^{(0)}$, which may have a significant effect on the computational performance of alm-DMET. More precisely, we observe that $D^{(0)}$ has to reflect the local (potentially fractional) number of particles in order to converge the algorithm (for more details see Appendix~\ref{app:NumTreatHub2d}). This together with the hyperparameter dependence indicates that as a non-convex optimization procedure, the convergence of alm-DMET may be more delicate than that of the semi-definite programming based approach in Ref.~\citenum{wu2020enhancing}}

We provide a pseudocode of the workflow for the full numerical procedure in Alg.~\ref{alg:ALM2}. The ALM convergence criterion (ALM-CC) consists of three numerical quantities that simultaneously need to reach pre-defined thresholds: (1) the Lagrange multiplier iteration step is smaller than $\tau_u$ (2) the 1-RDM iteration step (between $D^{(k+1)}$ and  $D^{(k)}$) is smaller than $\tau_D$, and (3) the {\it max norm}\footnote{The {\it max norm} is the element-wise $L_{p,q}$ matrix-norm with $p = q = \infty$, i.e. $ \|A\|_{\max} = \max_{ij} |a_{ij}|. $} of $\Delta$, where $\Delta={\rm diag} (\{D_x-P_x\})$, is smaller than $\tau_\Delta$.  In the subsequent numerical investigations we have set  $\tau_u= 10^{-6}$, $\tau_D= 10^{-8}$, and $\tau_\Delta= 10^{-6}$. The \ed{inner} iterations 
are performed until the iteration step reaches a pre-defined numerical threshold denoted $\tau_{D}^{\rm (pg)}$. Both \ed{outer} and \ed{inner} iterations are moreover capped with a maximal number of iterations. For the \ed{outer} iterations, we choose a maximum of 20000 iterations\ed{, and for the inner iterations, we choose a maximum of 5 iterations}. \ed{Note that the inner iterations are potentially computationally expensive since each step requires a full system diagonalization}; however, we have observed that for the first outer iterations a small number of inner iterations is in fact beneficial to the convergence of the ALM\ed{, see Appendix~\ref{app:mactroIts} exemplifying this for the 2D Hubbard model}. 
\ed{Comparing the numerical cost of the inner optimization with the cost of the optimization used in the global least-squares based DMET, we highlight that although the projected-gradient step requires the eigendecomposition of a potentially large matrix, it avoids the numerically more expensive computation of analytic gradients of the mean-field density matrix with respect to the correlation potential~\cite{wouters2016practical}, of which the cost is similar to that of density functional perturbation theory.}

\ed{Potential acceleration procedures like the direct inversion of the iterative subspace (DIIS)~\cite{Pulay1980} may also be applied to different optimization steps in the alm-DMET workflow, which are not used in this work.
}


\begin{algorithm}[H]
\label{alg:ALM2}
\caption{Correlation potential fitting using ALM}
 $k$ $ \leftarrow $ 0 (ALM iterator) \vspace{-2mm}\;
\While{ALM-CC is not reached}{
    \makebox[10mm]{\hfill$D$}  $\leftarrow $
        $D^{(k)}$\vspace{-2mm}\;
    \While{projected gradient convergence is not reached }{
        \makebox[10mm]{\hfill $\nabla_D \mathcal{L}_\alpha$ } $ \leftarrow $
            $f + u^{(k)} + \alpha^{(k)} \cdot \Delta$ \vspace{-2mm}\;
        \makebox[10mm]{\hfill $W$} $\leftarrow$
            $D - t^{(k)} \cdot \nabla \mathcal{L}_\alpha$\vspace{-2mm}\;
        \makebox[10mm]{\hfill $D$ } $ \leftarrow $ see Eq.~\eqref{eq:projGradStep},~\eqref{eqn:scalar_min}\vspace{-2mm}\;
        \makebox[10mm]{\hfill $\Delta$ } $ \leftarrow $
            ${\rm diag}(\{D_x -P_x\})$\vspace{-2mm}\;
        }
    \makebox[10mm]{\hfill$u^{(k+1)} $}  $\leftarrow $
        $u^{(k)} + \alpha^{(k)} \cdot \Delta$\vspace{-2mm}\;
    \makebox[10mm]{\hfill$D^{(k+1)} $}  $\leftarrow $
        $D$\vspace{-2mm}\;
    \makebox[10mm]{\hfill$\alpha^{(k+1)}$}  $ \leftarrow $ 
        $\frac{3}{2}\alpha^{(k+1)}$ \vspace{-2mm}\;
    \makebox[10mm]{\hfill$t^{(k+1)}$}  $ \leftarrow $ 
        $\frac{2}{3}t^{(k)}$ \vspace{-2mm}\;
    \makebox[10mm]{\hfill $k$ } $ \leftarrow $ $k+1$ \vspace{-2mm}\;
}
\end{algorithm}

\subsection{Postprocessing}
\label{sec:postprocessing}

At convergence of alm-DMET and when the exact matching condition is satisfied, the correlation potential is given by the Lagrange multiplier $u = \diag(\{u_x\})$. This can be seen by decomposing $D = CC^\dag$ 
and \ed{taking the derivative of $\mathcal{L}_\alpha$ with respect to ${C}$ (assuming real arithmetic for simplicity), i.e.}
\ed{
\[
\partial_C \mathcal{L}_\alpha(CC^\dagger, \{u_x\})
=
2(f+u)C,
\]
where $f$ is the Fock matrix.
Note that the penalty term vanishes when the matching condition is satisfied. Together with the orthogonality condition $C^{\dag} C=I$, at the stationary point of the augmented Lagrangian, the solution satisfies an eigenvalue equation
\begin{equation}
\label{eq:almHam}
H^\mathrm{ll} C = \Lambda  C, \quad H^\mathrm{ll} = f+ u.
\end{equation}
} 
We can therefore compute the low-level orbitals $\{\phi_m\}$ (the columns of $C$, corresponding to a Slater determinant $\Phi$, not necessarily of Aufbau occupancy) from the Hamiltonian $H^\mathrm{ll}$.

With the low-level orbitals at hand, we can check their occupation by computing $\Vert D \phi_m \Vert$. Given sufficiently well converged $u$, $\Vert D \phi_m \Vert$ will be equal to one for the occupied orbitals defining $D$, and zero for the unoccupied orbitals; recall that the idempotency condition in Eq.~\eqref{Eq:OptProb} ensures that the occupation of $\{\phi_m\}$ is either one or zero. This allows us to compute the occupation profile and to check the violation of the Aufbau principle. We emphasize that for a clear occupation profile, $u$ needs to be sufficiently well converged ($\tau_u \leq 10^{-6}$). \ed{If this threshold is not reached, the occupation profile might show values between zero and one (i.e. ``soft edges''). We emphasize that this does not indicate a fractional occupation, since alm-DMET yields an idempotent density by construction. In other words, the full convergence of $u$ is not necessary for the computational success of alm-DMET since $u$ is not utilized in the subsequent computations. Moreover we numerically observe that} the low-level 1-RDM converges faster than the Lagrange multiplier $u$, i.e. in our simulations the threshold $\tau_D$ was {\it always} reached before $\tau_u$ (such a behavior can be explained by Theorem~17.6 
in Ref.~\citenum{nocedal2006numerical}). 

\section{Numerical results}
\label{sec:numer}

We now focus on systems where SCF-DMET yields an artificially gapless low-level system and the assumption of NI-PS-V breaks down. We found this to be the case for a hole-doped 2D Hubbard model, a linear H$_{36}$ chain, and an H$_6$ model, but we emphasize that the appearance of a vanishing low-level gap is by no means restricted to said systems.

For comparison, we also consider two other algorithms. First, the standard least squares procedure to match the 1-RDMs via a correlation potential will be termed ls-DMET (we here use the global fitting implementation of ls-DMET~\cite{wouters2016practical}).
Second, the semi-definite programming formulation of the same fit will be referred to as cvx-DMET~\cite{wu2020enhancing}. The convex optimization is performed by the cvxpy package~\cite{diamond2016cvxpy, agrawal2018rewriting}. 
Note that when the low-level gap does not vanish, cvx-DMET guarantees the exact matching of density matrices under mild conditions, however, it may not return a meaningful result when the gap vanishes (the solver returns an exception) therefore we will only report results when an exact match is possible. 
All computations are performed using the \texttt{PySCF} software package~\cite{sun2018pyscf,sun2020recent}.



\subsection{2D Hubbard model}
\label{sec:2d_hub}
The first system we investigate is the 2D Hubbard model with periodic boundary conditions. In the case of strong electron-correlation effects, e.g. in the hole-doping regime with strong on-site interactions, the 2D Hubbard model represents a challenging problem in modern computational physics~\cite{zheng2017stripe,qin2021hubbard}. 
We here investigate a finite 2D Hubbard model on a six-by-six lattice. We begin with interaction strength $U = 8t$ for two different fillings; first, the less-problematic half-filling case with filling $n=1$, and second, the significantly more challenging hole-doped case with filling $n=\frac{32}{36}=\frac{8}{9}$.  In the hole-doped case, SCF-DMET can yield a vanishing low-level gap; moreover, the number of particles on the individual fragments is fractional, and we modify the initial guess of the 1-RDM in alm-DMET, i.e. $D^{(0)}$ in Sec.~\ref{sec:alm}, to reflect the local fractional number of particles in order to converge the algorithm, see Appendix~\ref{app:NumTreatHub2d} for a more detailed description.



For the alm-, cvx-, and ls-DMET computations we partition the system into nine fragments of size two-by-two, using the interacting bath formulation. The impurity is solved at the FCI level of theory in the spin-unrestricted formulation; the global high-level 1-RDM is obtained by democratic partitioning~\cite{wouters2016practical}. 
The initial 1-RDM for the DMET self-consistent cycles is obtained from an unrestricted Hartree-Fock (UHF) calculation. It is worth noting that the implementations of alm-, cvx- and ls-DMET aim to find the correlation potential on the global domain, i.e. they do not explicitly take advantage of the periodicity of the problem. However, we find in this system that the converged solutions are all cell-translationally invariant.


In the half-filling case, all algorithms find an exact match. Consequently, we find that {alm-,} {cvx-,} and ls-DMET achieve excellent agreement with each other \ed{and require the same number of DMET iterations}, see Table~\ref{tab:Hub_benchmark}. \ed{Regarding the outer iterations, we observe $n_{\rm outer}^{\rm(alm)} \approx 800$ 
for alm-DMET. Moreover, alm-DMET requires on average less than three inner iterations per outer iteration in order to converge, i.e. $\bar n_{\rm inner}^{\rm(alm)}\leq 3$. 
We emphasize, however, that by hyperparameter tuning the number of outer and inner iterations of alm-DMET can be significantly reduced such that $n_{\rm tot}^{\rm(alm)}= n_{\rm outer}^{\rm(alm)} \cdot \bar n_{\rm inner}^{\rm(alm)} \approx 100$, for more details see Appendix~\ref{app:mactroIts}. Note that this hyperparameter tuning does not affect the energy SCF trajectory presented in Table~\ref{tab:Hub_benchmark}.} 

\ed{In case of cvx-DMET, the number of optimization iterations per SCF iteration strongly depends on the 
convex optimization routine employed and its efficient implementation. Similarly, the number of optimization iterations per SCF iteration in ls-DMET depends on the solvers for the least square problems. We therefore do not report the number of outer iterations for cvx- and ls-DMET.}


\begin{table}[h!]
    \centering
    \begin{tabular}{c||c|c|c}
    It & alm-DMET & cvx-DMET & ls-DMET\\
    \hline
    1 & -0.52724    & -0.52724  & -0.52724 \\
    2 & -0.51731    & -0.51731  & -0.51731 \\
    3 & -0.51687    & -0.51687  & -0.51687 \\
    4 & -0.51685    & -0.51685  & -0.51685 \\
    5 & -0.51685    & -0.51685  & -0.51685 \\
    6 & -0.51685    & -0.51685  & -0.51685 \\
    \end{tabular}
    \caption{Comparison of convergence of different DMET algorithms during the self-consistent iterations. Tabulated is the energy per site (units of $t$) in the $6\times 6$ Hubbard model with $U=8.0t$ and periodic boundary conditions at half filling ($n=1$) as a function of the iteration number.}
    \label{tab:Hub_benchmark}
\end{table}

\begin{figure}[h!]
    \centering
    \includegraphics[width = \textwidth]{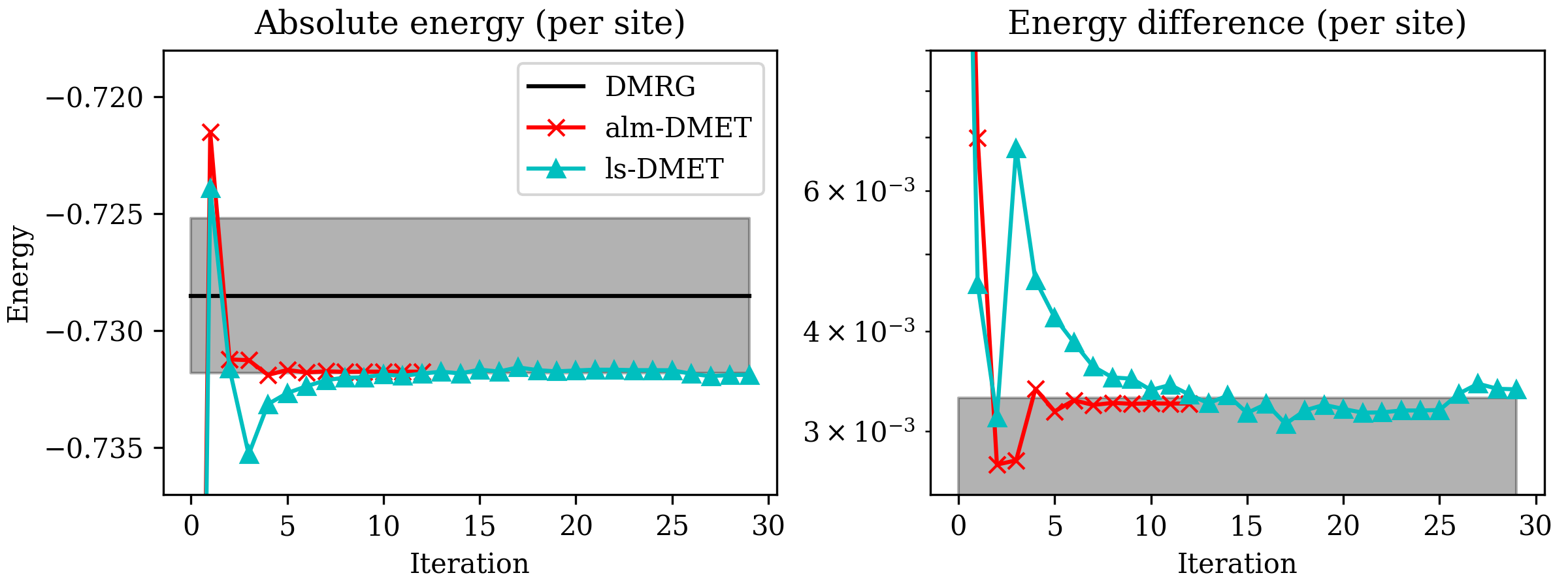}
    \caption{Energy per site (units of $t$) of the six-by-six 2D Hubbard model with $U = 8.0t$ and $n=32/36$. The shaded gray area around the DMRG solution corresponds to the DMRG extrapolation error $\Delta = 0.0033t$ (see Appendix~\ref{app:DMRG2DHub}).}
    \label{fig:Hub_its}
\end{figure}

Moving on to the hole-doped case, cvx-DMET and ls-DMET are not able to obtain an exact fit, which coincides with the appearance of a vanishing low-level gap during the DMET iterations in both algorithms. We do not report the cvx-DMET results when the constraints are not satisfied as the cvxpy solver returns an arbitrary result. In the case of ls-DMET \ed{(see Fig.~\ref{fig:Hub_its})}, the algorithm converges slowly with respect to the self-consistent iterations 
to an inexact match,
where the discrepancy in the low-level and high-level 1-RDM blocks is in the range of $0.01$--$0.1$ in Frobenius norm per site, and $0.01$--$0.1$ in max norm. \ed{Hence, the optimization of the least squares problem does not converge to the threshold $\tau_{\Delta}$.} 
(We note that the energy difference between iterations does not converge to the set threshold of $10^{-6}t$ within 30 iterations). 
We note that even standard Hartree-Fock calculations for the hole-doped case are difficult to converge and we therefore need to use finite temperature smearing for the very first mean-field calculation (we here used the inverse temperature $\beta = 100$). We emphasize however, that after this first mean-field calculation, all subsequent computations in alm-, and ls-DMET are performed in the zero temperature limit.

\begin{figure}[h!]
    \centering
    \includegraphics[width = 0.55\textwidth]{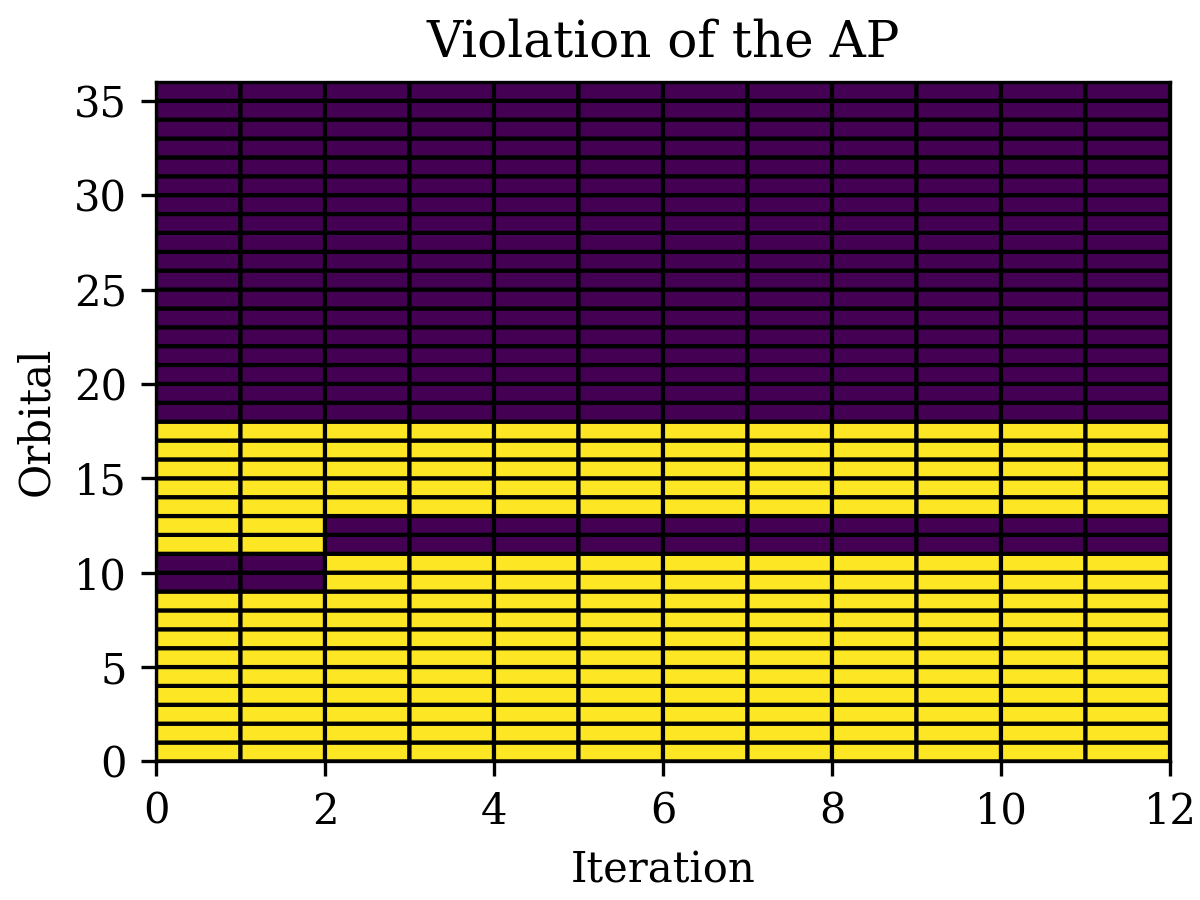}
    \caption{Visualization of the violation of Aufbau principle in alm-DMET in the 2D Hubbard model. The Y-axis describes the orbitals sorted in increasing order by the corresponding orbital energy; the X-axis is the iteration number, and the yellow squares correspond to the occupied orbitals. The Aufbau violating profile can be seen by the presence of dark entries in the middle of the yellow blocks. }
    \label{fig:Hub_AP}
\end{figure}

In contrast to ls-DMET, alm-DMET yields an exact match at every iteration of the DMET self-consistent cycle, with a discrepancy of about $10^{-7}$ in both the Frobenius and max norm. 
The match is achieved by violating the Aufbau principle (see Fig.~\ref{fig:Hub_AP}), as determined by the procedure described in Sec.~\ref{sec:alm}.
Recall that in the hole-doped case ($n=\frac{32}{36}$) in the spin-unrestricted formulation, there are $16$ electrons per spin component. At convergence, we find that the occupation profile of alm-DMET creates two holes (i.e.~orbitals $12$ and $13$ are unoccupied) below the Fermi surface. Furthermore, the convergence of the overall DMET SCF iterations is faster and smoother using alm-DMET. Compared to the extrapolated DMRG ``exact" result for this system, 
the energies obtained by alm-DMET and ls-DMET \ed{(see Fig.~\ref{fig:Hub_its})} using the small $2\times 2$ impurities are quite close to the exact result (within the DMRG extrapolation error).
\ed{We observe that alm requires 2000--3000 outer iterations with less than three inner iterations.} 

\begin{figure}[h!]
    \centering
    \includegraphics[width = 0.5\textwidth]{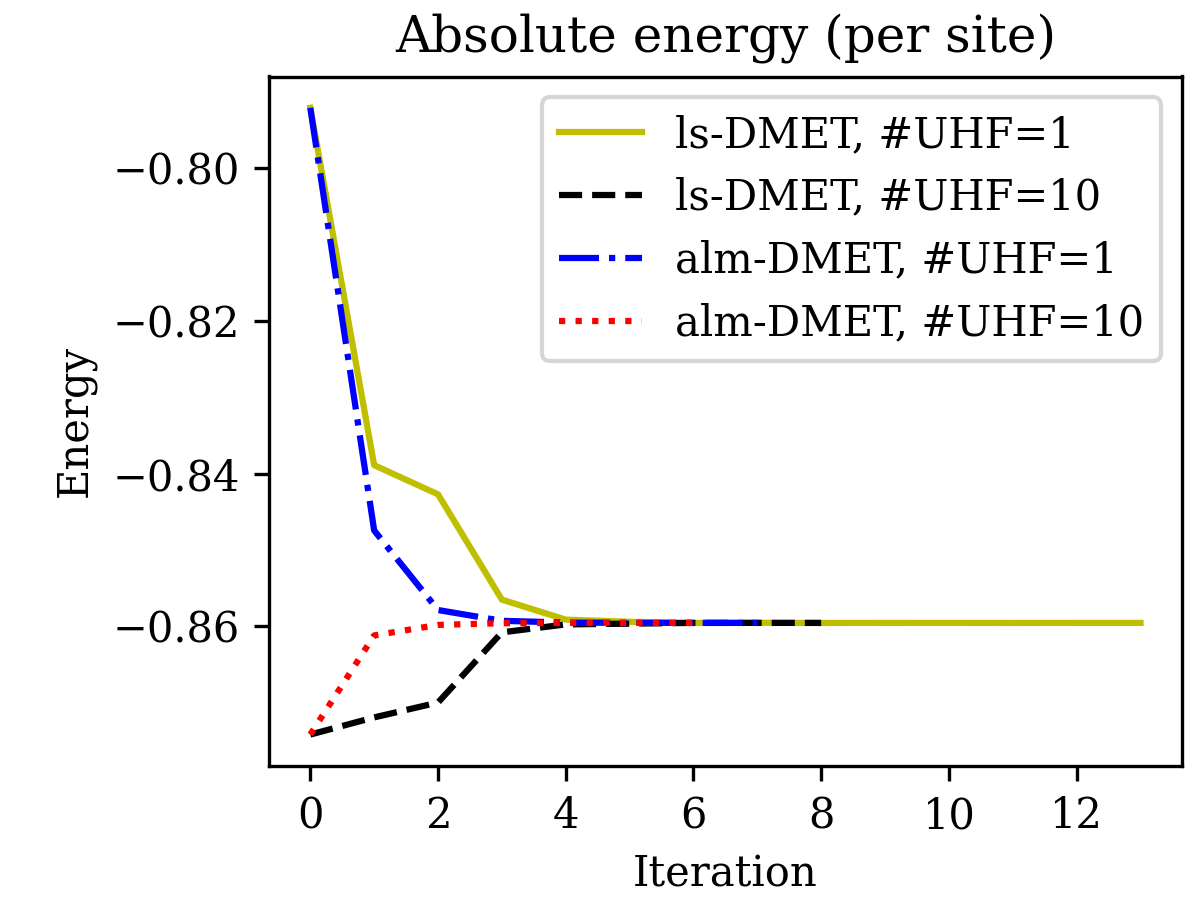}
    \caption{Convergence of the energy per site of the six-by-six 2D Hubbard model with $U = 4.0t$, and $n = 1.0$ using alm-DMET
    for different initial guesses corresponding to one and ten UHF iterations, respectively.}
    \label{fig:Hub_init_dep}
\end{figure}

We finish our investigation of the 2D Hubbard model by investigating the initial state dependence of the alm-DMET self-consistency. We perform a similar study as in Ref.~\citenum{wu2019projected}, where the initial 1-RDM for alm-DMET 
is taken from unconverged UHF mean-field solutions after performing one and ten Hartree-Fock iterations, respectively.
We observe that for the 2D Hubbard model with six-by-six sites and on-site interaction strength $U = 4.0t$ at half filling ($ n = 1.0$), alm-DMET 
is independent of the initial-state guess, see Fig.~\ref{fig:Hub_init_dep}. 
We emphasize that  Fig.~\ref{fig:Hub_init_dep} reports the initial-state independence of the alm-DMET self-consistent cycle, which needs to be distinguished from the initial 1-RDM guess $D^{(0)}$ used in the ALM procedure, see Alg.~\ref{alg:ALM2} and Appendix ~\ref{app:NumTreatHub2d}. 
The latter can affect the convergence of the alm-DMET optimization as well as the self-consistent DMET convergence.
  
\subsection{Hydrogen chain H$_{36}$}
\label{sec:H36}

A DMET calculation should systematically converge to the solution of the full system when the fragment size is continuously increased. Indeed, if there is merely one fragment containing the full system, the result from DMET is as accurate as the global high-level solution. However, it has been reported that the fragment size cannot be arbitrarily increased without running into a gapless low-level system~\cite{wu2020enhancing}. When using the cvx-DMET procedure, one may obtain some arbitrary results as discussed in Ref.~\citenum{wu2020enhancing}. In some of these cases, ls-DMET can be relatively more stable via an inexact match, however, this is not entirely satisfactory as it raises the issue of non-unique solutions.
Here we use the linear H$_{36}$ test system discussed in Ref.~\citenum{wu2020enhancing} to study this problem with alm-DMET. The H-H bond length is chosen to be 1 \AA, we use the STO-6G basis set, and we solve the high-level problem at the CCSD level of theory, with the initial 1-RDM  obtained from a restricted Hartree-Fock (RHF) calculation. 
The orthogonal basis in the DMET calculations is the symmetrically orthogonalized AO basis.

\begin{figure}[h!]
    \centering
    \includegraphics[width= \textwidth]{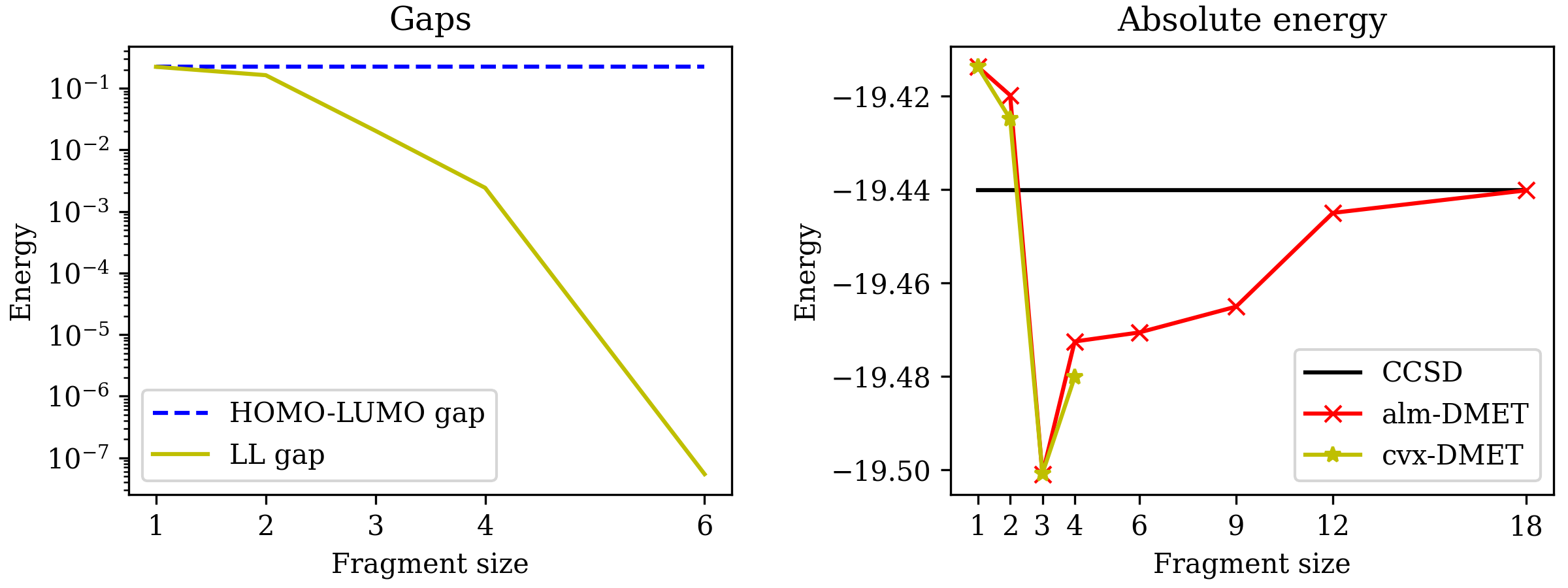}
    \caption{Comparison of the low-level gap and the RHF HOMO-LUMO gap (left) and the total energy  (right) of the linear H$_{36}$ system for different fragment sizes.}
    \label{fig:H36_pes}
\end{figure}

\begin{figure}
    \centering
    \includegraphics[width= 0.5 \textwidth]{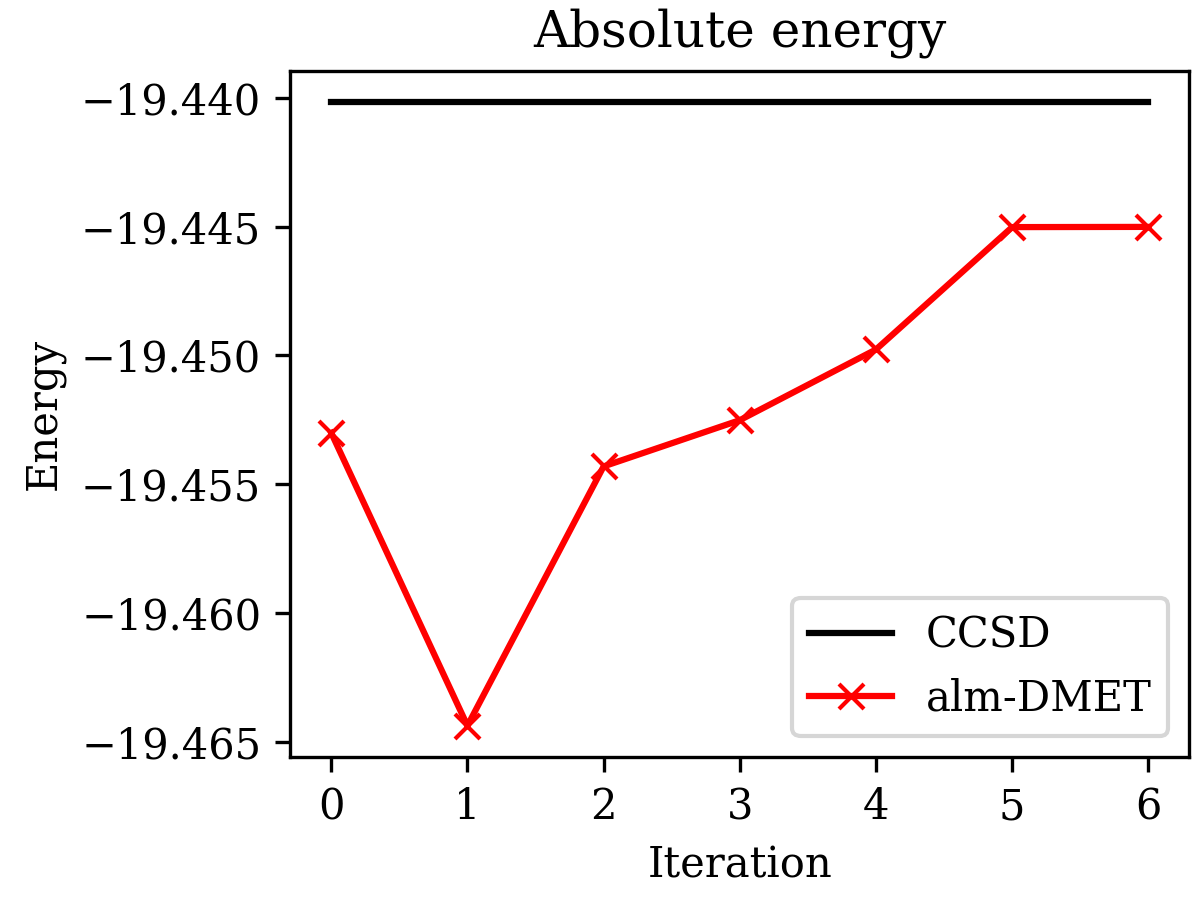}
    \caption{Self-consistent iterations of alm-DMET for linear H$_{36}$ with fragment size equal to 12. Total energy compared to the CCSD benchmark energy.}
    \label{fig:h36_its}
\end{figure}

We first confirm that cvx-DMET rapidly yields an artificially gapless low-level system as we increase the fragment size, see Fig.~\ref{fig:H36_pes} (left). This causes cvx-DMET to fail for fragment sizes of 6 or larger. Simulating the system with alm-DMET, we obtain stable solutions for all fragment sizes, see Fig.~\ref{fig:H36_pes} (right). Moreover, we observe the systematic convergence of the alm-DMET energies towards the high-accuracy solution of the full system: for fragment size 18 the energy difference between the high-level solution of the full system and alm-DMET is approximately $10^{-8}$ Hartree. The SCF procedure using alm-DMET also converged within a reasonable number of iterations ($\leq 15$) for all fragment sizes reported in Fig.~\ref{fig:H36_pes}. We illustrate the convergence for fragment size 12 in Fig.~\ref{fig:h36_its}.

\begin{figure}[h!]
    \centering
    \includegraphics[width = \textwidth]{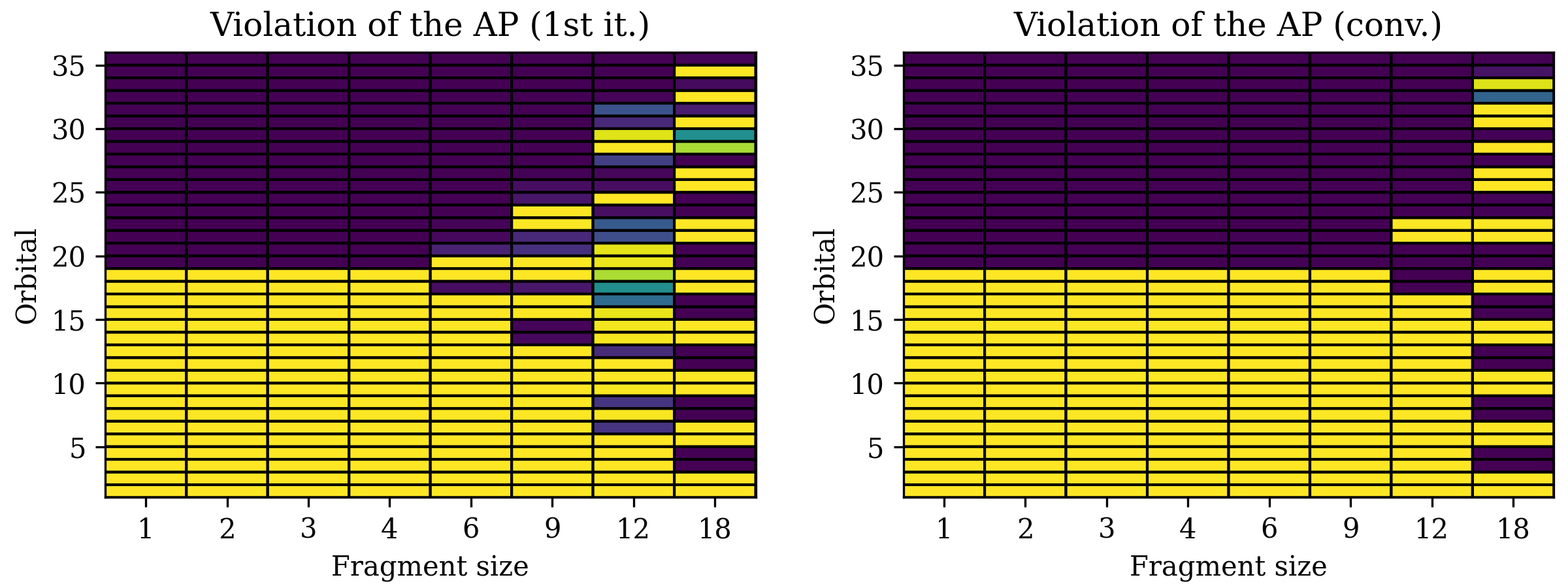}
    \caption{Visualization of the violation of the Aufbau principle in the linear H$_{36}$ system for the first (left) and converged alm-DMET iterations (right). The Y-axis describes the orbitals sorted in increasing order by the corresponding orbital energy; the X-axis is the fragment size, and the yellow squares correspond to the occupied orbitals.}
    \label{fig:H36_viol_AP}
\end{figure}

In order to achieve a high-accuracy fit of the 1-RDM for fragment size equal to 6 or larger alm-DMET violates the Aufbau principle during the DMET self-consistent iterations, see Fig.~\ref{fig:H36_viol_AP}. 
Note that, for fragment sizes 
$6$ and $9$, the violation of the Aufbau principle that occurs in the first SCF iteration is removed in the final converged 1-RDMs. 
However, for fragment sizes 12 or 18 the Aufbau principle is violated even in the converged 1-RDMs. We emphasize that for larger fragment sizes it becomes more difficult 
to fully converge the Lagrange multipliers in alm-DMET, even though the density matrix matching constraint is well converged; this is related to a known property of the ALM, see discussion at the end of Sec.~\ref{sec:postprocessing} \ed{(and Theorem~17.6 in Ref.~\citenum{nocedal2006numerical})}. This reduces the accuracy of the test for the Aufbau principle since the correlation potential $u$ is not fully determined, leading to \ed{an occupation profile with ``soft edges'' (i.e. seemingly fractional occupancies)} in Fig.~\ref{fig:H36_viol_AP} for fragment sizes 12 and 18. \ed{We emphasize, however, that the density matrix matching constraints are well converged, i.e. $\tau_D$ and $\tau_\Delta$ are reached, and that the approximate density is idempotent---it is merely the insufficiently converged Lagrange multiplier that yields soft edges in the occupation profile (with respect to the matrix $f+u$).}

\ed{
For the fragment sizes of 1--4, alm-DMET requires 4--6 DMET-iterations with 2000--3000 outer iterations and less than three inner iterations on average. For the fragment sizes 6, 9, 12, and 18 alm-DMET requires over 10000 outer iterations with less than three inner iterations on average. Again, the chosen hyperparameters of the alm-DMET procedure may not yield the fastest convergence possible for the considered system (similar to the 2D Hubbard example, see Appendix~\ref{app:mactroIts}). Based on our experiences with the 2D Hubbard model, we believe that further fine tuning of the hyperparameters can significantly reduce the number of outer and inner iterations of alm-DMET. 
}




\subsection{H$_6$ model}
\label{sec:H6}

\begin{figure}
    \centering
    \includegraphics[width = 0.9\textwidth]{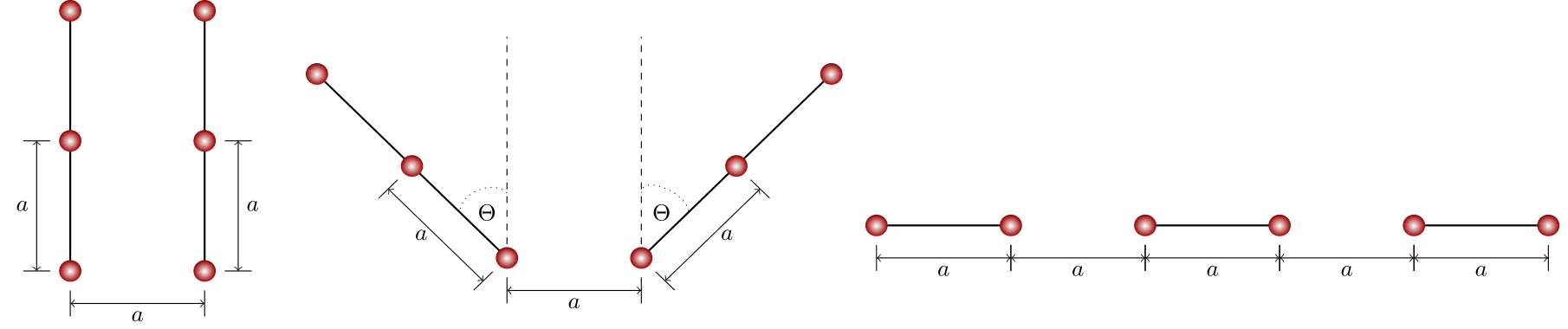}
    \caption{Nuclear configuration for the H$_6$ chain undergoing the transition from a rectangular geometry (left) to a linear geometry (right).}
    \label{fig:H6_model}
\end{figure}

Finally, we study an extension of the H$_4$ model~\cite{jankowski1980applicability} to 6 hydrogen atoms, which we shall refer to as the {H$_6$ model}. We consider six hydrogen atoms undergoing a similar transition as the H$_4$ model, i.e. from a rectangular configuration to a linear configuration. This transition is modelled by the system parameter $\Theta$, where $\Theta= 0$ corresponds to the rectangular geometry and $\Theta=\pi/2$ is the linear geometry, respectively (see Fig.~\ref{fig:H6_model}). 


We use a (non-minimal) 6-31++G basis set, and we choose the bond length between the hydrogen atoms to be 1 \AA~(i.e. $a=1$ \AA~in Fig.~\ref{fig:H6_model}). We partition the system into six fragments consisting of one atom each. The initial 1-RDM is obtained from RHF calculations, and the impurity system is solved at the FCI level of theory. The orthogonal basis in the DMET calculations is generated by the meta-L\"owdin localization method~\cite{Sun14qmmm}.


\begin{figure}
    \centering
    \includegraphics[width=0.5\textwidth]{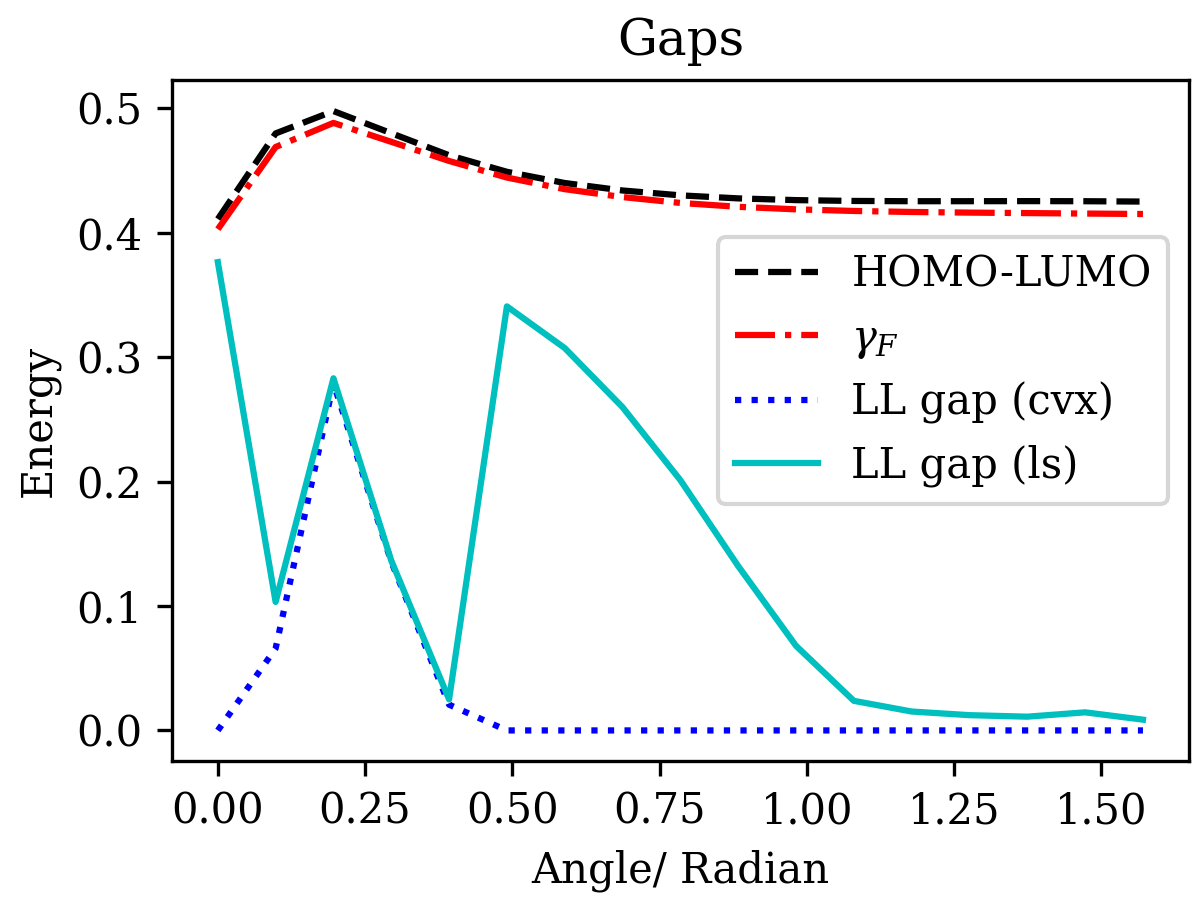}
    \caption{Comparison of the cvx-DMET low-level gap (LL gap (cvx)), the ls-DMET low-level gap (LL gap (ls)), the fundamental gap ($\gamma_{\rm F}$), and the RHF HOMO-LUMO gap (HOMO-LUMO) as a function of the system parameter $\Theta$. The fundamental gap is computed using FCI. 
    }
    \label{fig:LL-gap-H6}
\end{figure}

We compare the low-level gap obtained in cvx- and ls-DMET with 
the fundamental gap (i.e. $\gamma_{\rm F} = E_{\rm IP} - E_{\rm AP}$ where $E_{\rm IP}$ and $E_{\rm EA}$ are the ionization potential and electron affinity, respectively) obtained from FCI, see Fig.~\ref{fig:LL-gap-H6}. 

\begin{figure}
    \centering
    \includegraphics[width= \textwidth]{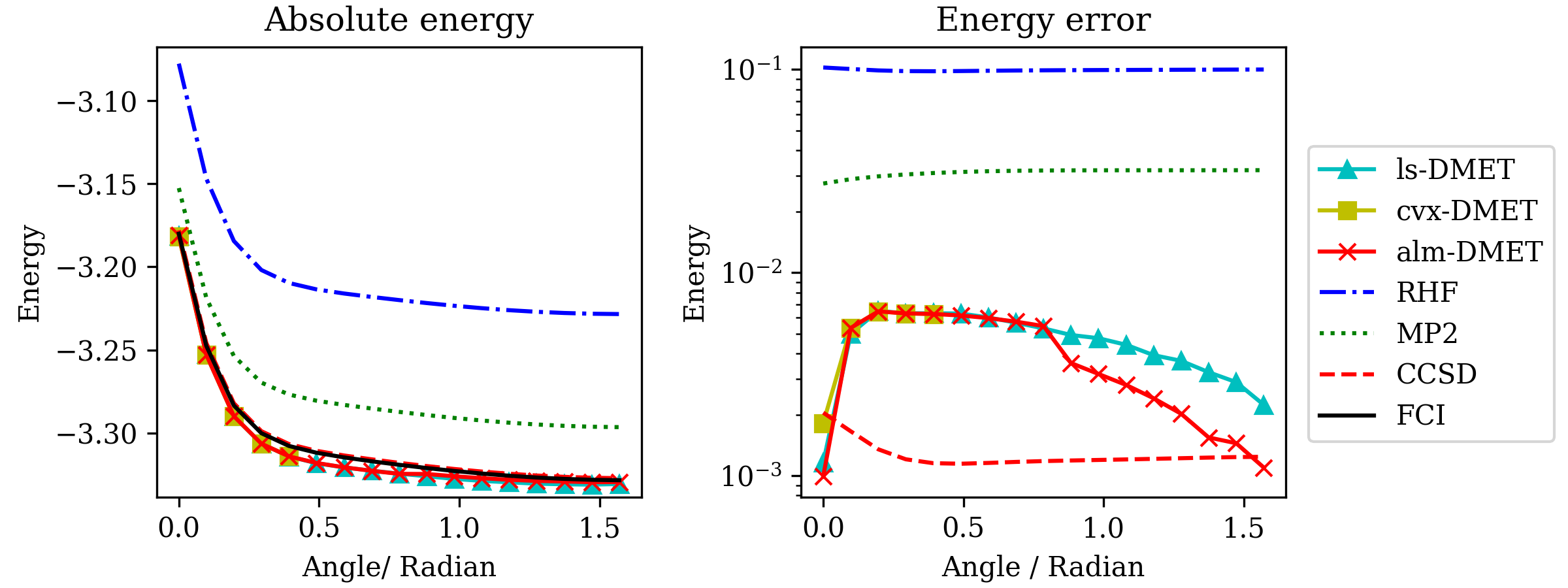}
    \caption{(Left) Potential energy surface of the H$_6$ model as a function of the system parameter $\Theta$ using the 6-31++G basis. (Right) Energy error compared to the FCI solution as a function of $\Theta$ (errors from other methods shown for scale).
    }
    \label{fig:H6_pes}
\end{figure}

As confirmed here, the vanishing low-level gap reflects the fictitious nature of the low-level DMET Hamiltonian, and thus there is no obvious connection between the vanishing low-level gap in DMET and the many-body gap. When the cvx-DMET gap is finite, the ls-DMET gap and cvx-DMET gap agree fairly well. However, when the cvx-DMET gap vanishes (indicating no exact matching can be found), the ls-DMET gap is somewhat different indicating that the ls-DMET is producing an inexact match. Note that when cvx-DMET and ls-DMET both find an exact match, their energies are  
identical up to chemical accuracy, see Fig.~\ref{fig:H6_pes}.
\ed{The alm-DMET approach circumvents the numerical problems of a vanishing low-level gap by directly fitting the density matrix while not enforcing the Aufbau principle. A consequence of the violation of the Aufbau principle is that a low-level gap is not well-defined. Moreover, the spectral gaps of the low-level Hamiltonian appear to have no direct importance for the numerical success of alm-DMET, opposed to ls- and cvx-DMET.}

In the gapless regions where cvx-DMET fails to converge and ls-DMET produces an inexact match,  alm-DMET still finds an exact match. The absolute error of the various DMET energies is shown in Fig.~\ref{fig:H6_pes}, with errors from CCSD, MP2 (second-order M{\o}ller-Plesset perturbation theory), and RHF shown for scale. 

\begin{figure}
    \centering
    \includegraphics[width = \textwidth]{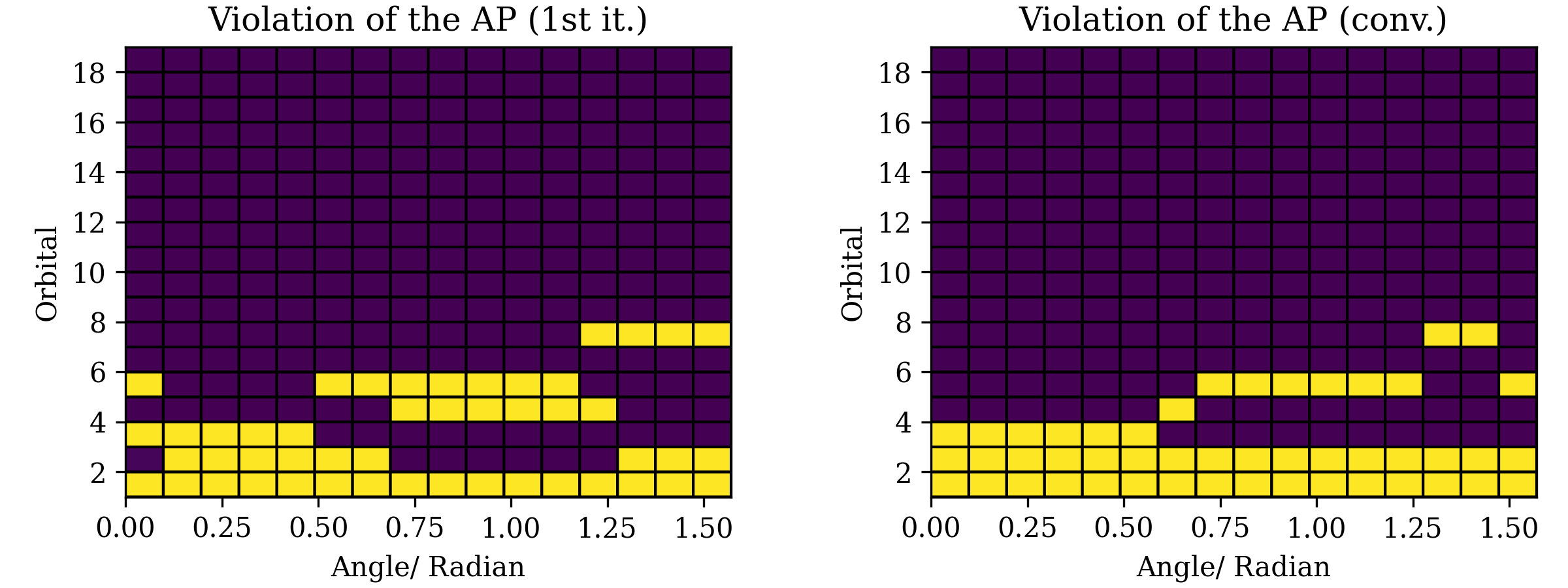}
    \caption{Visualization of the violation of the Aufbau principle in the H$_6$ model for the first (left) and the converged alm-DMET iterations (right). The Y-axis describes the orbitals sorted in increasing order by the corresponding orbital energy; the X-axis describes the system parameter $\Theta$, and the yellow entries correspond to the occupied orbitals.
    }
    \label{fig:H6_viol_AP}
\end{figure}

For $\Theta \geq 0.5$, alm-DMET violates the Aufbau principle during the self-consistent iterations, see Fig.~\ref{fig:H6_viol_AP}. 
Similarly to the H$_{36}$ linear model, we observe that the violation of the Aufbau principle in the first SCF iteration for some system parameter $\Theta$ does not imply that it is violated in the final converged density matrix (see the comparison between the left and right panels in Fig.~\ref{fig:H6_viol_AP}). From Fig.~\ref{fig:H6_pes}, we see that the exact matching achieved by alm-DMET leads to an improved energy error relative to the global FCI solution compared to ls-DMET,  although the violation of the Aufbau principle in alm-DMET creates a small discontinuity in the error plot.

\begin{figure}
    \centering
    \includegraphics[width = \textwidth]{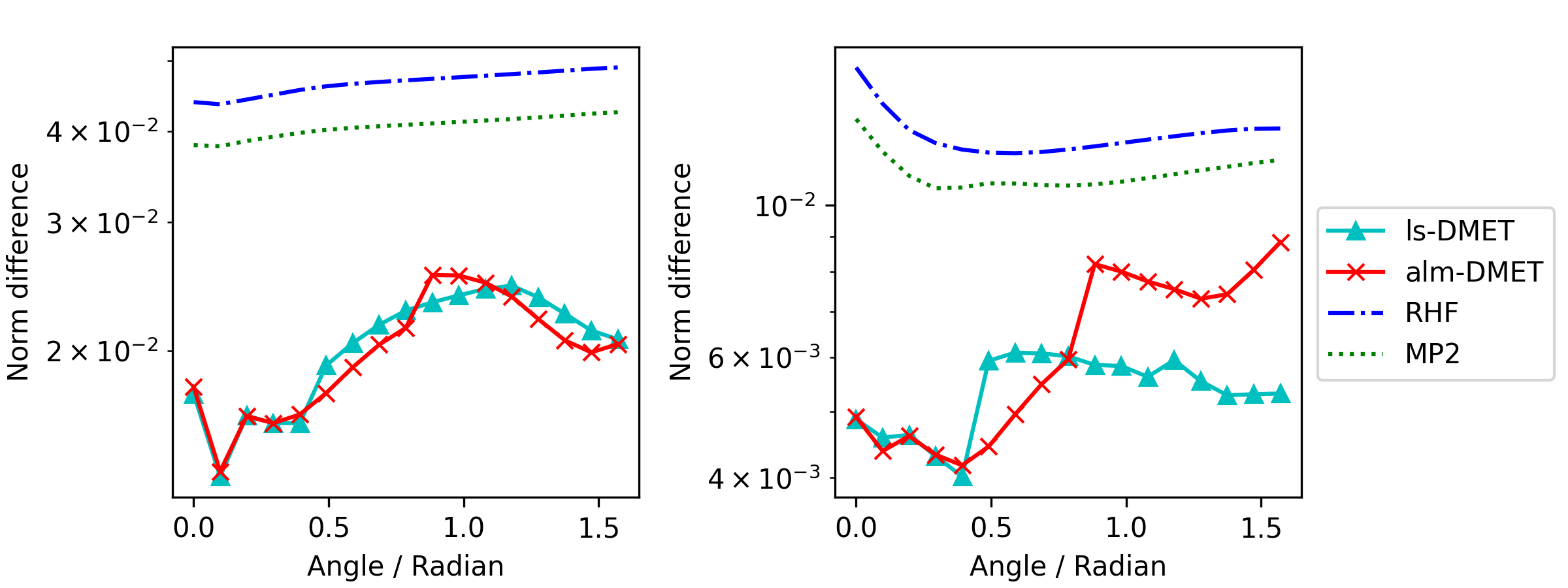}
    \caption{Difference (restricted to the impurity blocks), of the RHF, MP2, alm-DMET, and ls-DMET 1-RDM blocks compared to the FCI 1-RDM using the Frobenius norm (left) and the max norm (right). 
    }
    \label{fig:Fit_acc-H6}
\end{figure}

Comparing the accuracy of the density matrices (restricted to the impurity blocks) at different levels of theory, we observe that alm- and ls-DMET yield much better approximations of the 1-RDM to the FCI benchmark than that of RHF or MP2 (see Fig.~\ref{fig:Fit_acc-H6}). 
\ed{
In order to reach convergence, the alm-DMET requires 4000--12000 outer iterations with less than three inner iterations on average.   
}



\section{Conclusion}

In this paper, we carefully examined the non-interacting pure-state $v$-representability (NI-PS-V) assumption that underlies current SCF-DMET calculations. 
We presented numerical evidence showing that this assumption is violated in common scenarios, which leads to gapless low-level Hamiltonians and high-level 1-RDM blocks that cannot be exactly matched (cf. Proposition 2 in Ref.~\citenum{wu2020enhancing}). We proposed an alternative and more direct global fitting procedure of the high-level 1-RDMs, which is based on an augmented Lagrangian formulation (alm-DMET). The alm-DMET method relaxes the NI-PS-V assumption, which allows the pure state to follow any occupation profile---possibly violating the Aufbau principle---while yielding an idempotent low-level 1-RDM. 

We numerically tested the alm-DMET method for three prototypical systems which we identify as displaying violations of NI-PS-V: a hole-doped 2D Hubbard model, a linear H$_{36}$ system, and an H$_6$ model.
In each case, relaxing the Aufbau principle allows the alm-DMET method to satisfy the matching condition exactly. We find that this improves the numerical accuracy of the embedding. In the case of H$_{36}$ this allows us to demonstrate the systematic convergence to the exact global result in the large fragment limit while retaining the exact matching condition, which was not previously possible in Ref.~\citenum{wu2020enhancing}.

The ability to achieve exact matching 1-RDMs addresses an important issue of the SCF-DMET procedures since its inception. However, the augmented Lagrangian formulation is not a convex optimization problem, and is expected to be less robust compared to the SDP reformulation~\cite{wu2020enhancing} of the correlation potential fitting problem when NI-PS-V holds. \ed{For instance, the performance of alm-DMET can depend on the initial density $D^{(0)}$ as well as the  hyperparameters.}
A more severe problem caused by the relaxation of the Aufbau principle is that the occupation profile can change discontinuously along a continuous reaction coordinate in the absence of bond breaking or phase transitions.
As seen in our calculations on the H$_6$ model, this can lead to discontinuous potential energy surfaces, which can impede the practical application of the alm-DMET method.
One potential remedy is to further relax the NI-PS-V condition to allow the low-level density matrix to be non-idempotent, with violations of idempotency occurring in a continuous manner. However, the number of bath orbitals may then need to be chosen to be larger than the fragment size, which may potentially reduce the computational efficiency of DMET.
Furthermore, the number of bath orbitals still needs to be judiciously chosen to avoid introducing another source of discontinuity on the potential energy surface.
Finally, we expect that the potential violation of the NI-PS-V condition may be a general feature in quantum embedding theories beyond DMET.

\section*{Acknowledgement}

This work was partially supported by the Air Force Office of Scientific Research under award number FA9550-18-1-0095 (G.K.C., L.L.), by the Department of Energy under Grant No. DE-SC0017867 (F.M.F., R.K.), and under Grant No. DE-SC0018140 (Z.-H. C.). Z.-H. C. acknowledges support from the Eddleman Quantum Institute through a graduate research fellowship. G.K.C.  and L.L. are Simons Investigators.



\appendix

\section{Treatment of fractional occupancy in the 2D Hubbard model}
\label{app:NumTreatHub2d}



\begin{figure}
    \centering
    \includegraphics{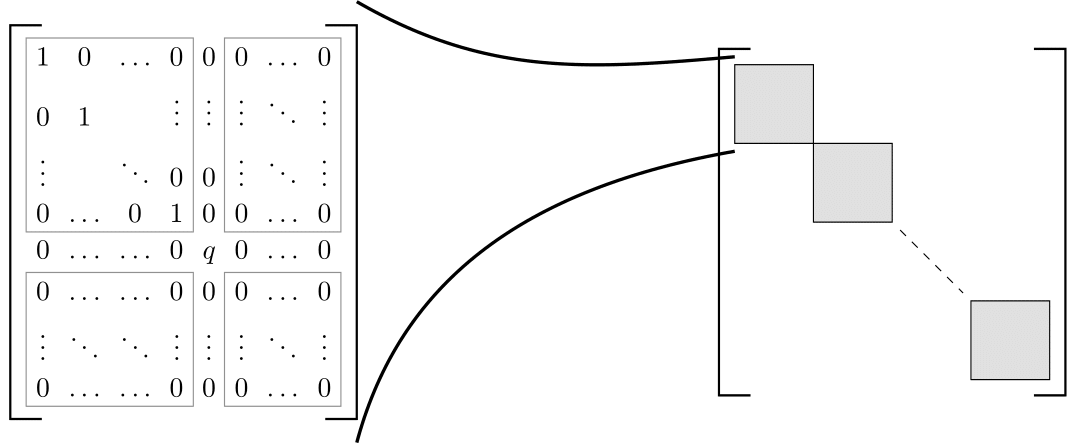}
    \caption{Schematic representation of the block-diagonal structure of $D^{(0)}$ and a depiction of a matrix block corresponding to one fragment. The value $q=n_{\rm occ}^{(x)}- \lfloor n_{\rm occ}^{(x)} \rfloor$.}
    \label{fig:D0}
\end{figure}

In Sec.~\ref{sec:numer} we investigate the two-dimensional hole-doped Hubbard model with six-by-six sites (periodic boundary conditions),  divided into nine two-by-two fragments with a total number of 32 electrons yielding $16/9\approx 1.78$ spin-up and spin-down electrons per fragment. We find that in order to converge,
alm-DMET needs to start from a properly chosen initial 1-RDM $D^{(0)}$ reflecting the number of electrons per fragment. We construct $D^{(0)}$ as a diagonal matrix, where the diagonal elements corresponding to the individual impurities represent the (fractional) number of particles in the fragment, $n_{\rm occ}^{(x)}$. 
To that end, we set the first $\lfloor n_{\rm occ}^{(x)} \rfloor$ entries to be one, the following entry to be $n_{\rm occ}^{(x)}- \lfloor n_{\rm occ}^{(x)} \rfloor$, and the remaining entries to be zero, see Fig.~\ref{fig:D0}. 

\section{\ed{Outer and inner iterations in the 2D Hubbard model}}
\label{app:mactroIts}

The fitting of the high-level density matrix requires to solve an optimization problem in each SCF iteration. In the case of alm-DMET we refer to this as the {\it outer optimization}. As mentioned in Section~\ref{sec:numer}, in case of cvx-, and ls-DMET, the number of optimization iterations per SCF iteration strongly depends on the optimization routine employed and its efficient implementation---a direct comparison seems therefore uninsightful. 
Subsequently, we elaborate on the outer and inner optimization performed in alm-DMET exemplified by the 2D Hubbard model. First, we numerically investigate the effect of the hyperparameter configuration on the number of outer iterations. Second, we investigate the application of an {\it ad hoc} bound in the inner optimization (i.e. to the projected gradient iterations).

\begin{figure}
    \centering
    \includegraphics[width= 0.8\textwidth]{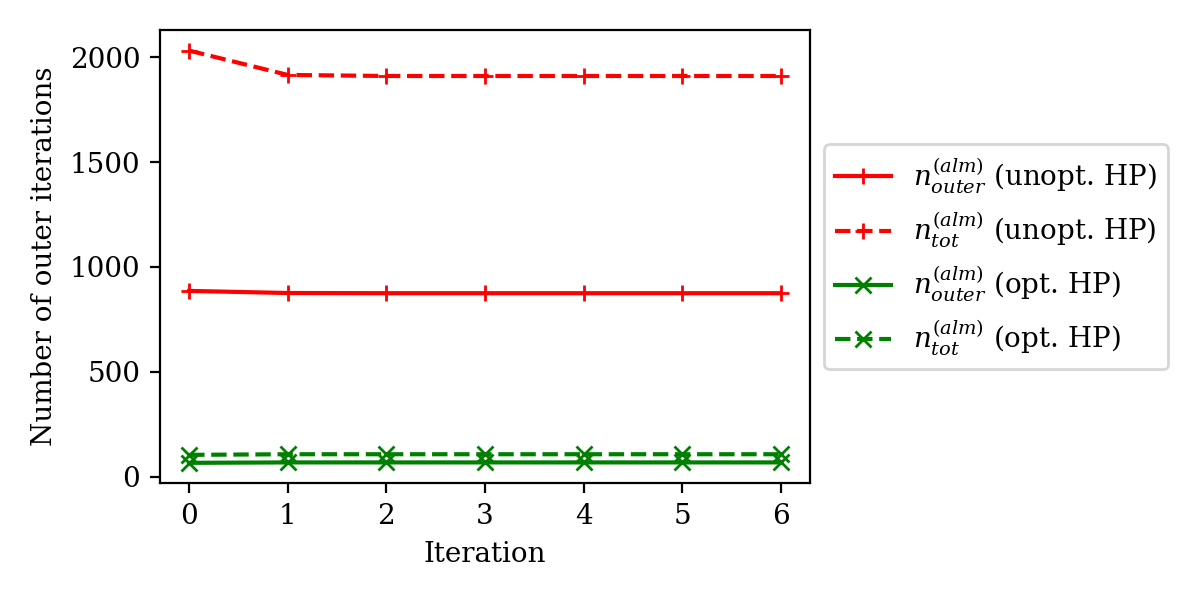}
    \caption{Comparison of the number of outer iterations of alm-, and ls-DMET for an optimized hyperparameter configuration (opt. HP) and an unoptimized hyperparameter configuration (unopt. HP) defined in Section~\ref{sec:alm}. 
    }
    \label{fig:Macro_its}
\end{figure}

In Fig.~\ref{fig:Macro_its} we report the number of outer iterations for alm-DMET, i.e. $n_{\rm  outer}^{\rm (alm)}$. Since each outer iteration requires to solve an additional optimization procedure, Fig.~\ref{fig:Macro_its} also includes the product of the average number of inner iterations with the number of outer iterations, i.e. $n_{\rm tot}^{\rm (alm)} = \bar{n}_{\rm  inner}^{\rm (alm)} \cdot n_{\rm  outer}^{\rm (alm)}$. The latter can be used to reflect the number of full system diagonalizations needed in alm-DMET per SCF iteration. By adjusting the hyperparameters (see the {\it opt.~HP} graphs in Fig.~\ref{fig:Macro_its}) the number of optimization steps can be significantly reduced. We set $t_{\rm init} = 0.6$, $t_{\rm min}= 0.001$, $\alpha_0= 5$, $\alpha_{\rm max} = 19$, update the hyperparameters every 10 outer iterations and bound the number of inner iterations by two. Note that opposed to the (unoptimized) hyperparameter setting presented in  Section~\ref{sec:alm}, the above hyperparameter configuration will not yield optimal convergence (or convergence at all) for all systems presented in Section~\ref{sec:numer}. This shows that the numerical performance of alm-DMET is highly dependent on the hyperparameter configuration indicating that the presented method can be further improved to yield better and more robust numerical performance. Meanwhile the alm-DMET trajectories for the optimized and unoptimized hyperparameter configurations are identical in energy for each SCF iteration yielding the results presented in Table~\ref{tab:Hub_benchmark}.

\begin{figure}
    \centering
    \includegraphics[width=0.7\textwidth]{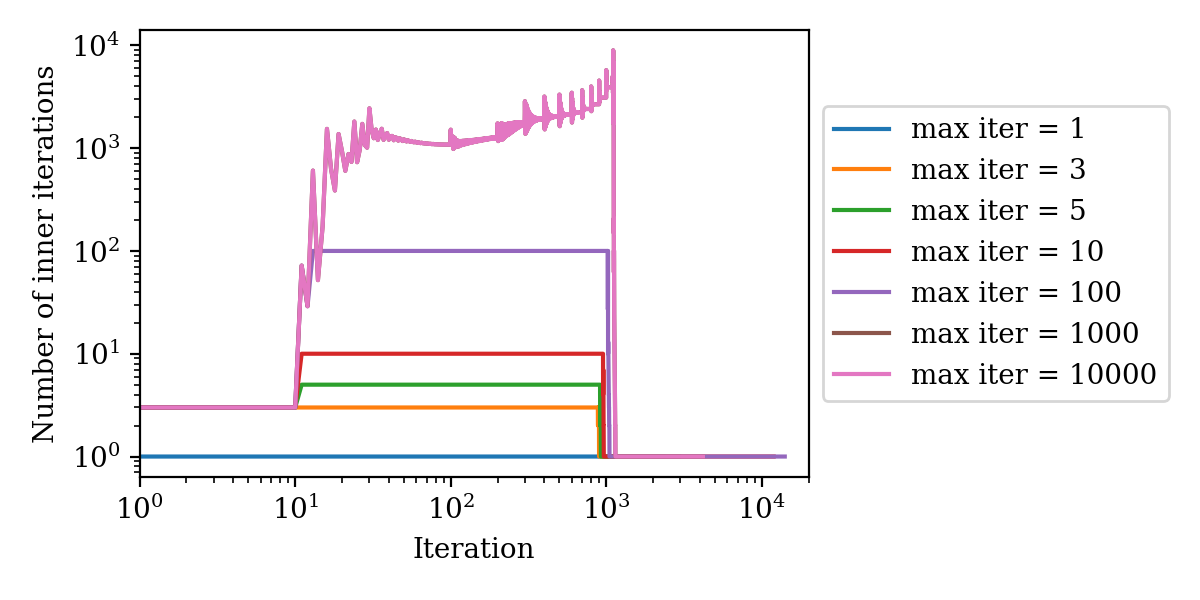}
    \caption{The number of inner iterations per outer iteration for the first DMET-iteration of the hole-doped 2D Hubbard model ($n=\frac{32}{36}$) for different maximal numbers of inner iterations (max iter).}
    \label{fig:HubMics}
\end{figure}

Employing a bound on the inner iterations limits the number of diagonalization operations, and reduces the cost of the alm procedure. In Fig.~\ref{fig:HubMics} we report the number of inner iterations per outer iteration for the first DMET-iteration of the hole-doped 2D Hubbard model ($n=\frac{32}{36}$). We observe that after 10 iterations with a small penalty parameter ($\alpha=0.1$) the projected gradient procedure can require a large amount of iterations to reach the convergence threshold $\tau_D^{\rm pg}=10^{-4}$. After approximately 1000 iterations the procedure appears to have stabilized and already one projected gradient step suffices to reach the convergence threshold $\tau_D^{\rm pg}$. 

Note that enlarging the maximal number of inner iterations not only increases the number of full system diagonalizations that are performed, but we also observe an increase in the number of outer iterations, see Table~\ref{tab:Macs_of_mics}, which further increases the total numerical costs. 

\begin{table}[h!]
    \centering
    \begin{tabular}{c|ccccccc}
         Bound of inner iterations &  1 & 3 & 5 & 10 & 100 & 1000 & 10000\\
         \hline
         Number of outer iterations & 2946 & 2910 & 2336 & 11928 & 14053 & 3981 & 4185\\
    \end{tabular}
    \caption{Number of outer iterations in dependence of the enforced bound of inner iterations.}
    \label{tab:Macs_of_mics}
\end{table}

We believe that the efficiency of our algorithm can be significantly improved if there is a systematic way to robustly set the hyperparameters. 
The results of this section illustrate the importance of setting hyperparameters in alm-DMET, and provide empirical evidence that within the current optimization algorithm, it is beneficial to limit the number of projected gradient steps.

\section{DMRG benchmark of the 2D Hubbard model}
\label{app:DMRG2DHub}

\begin{figure}
    \centering
    \includegraphics[width=0.5\textwidth]{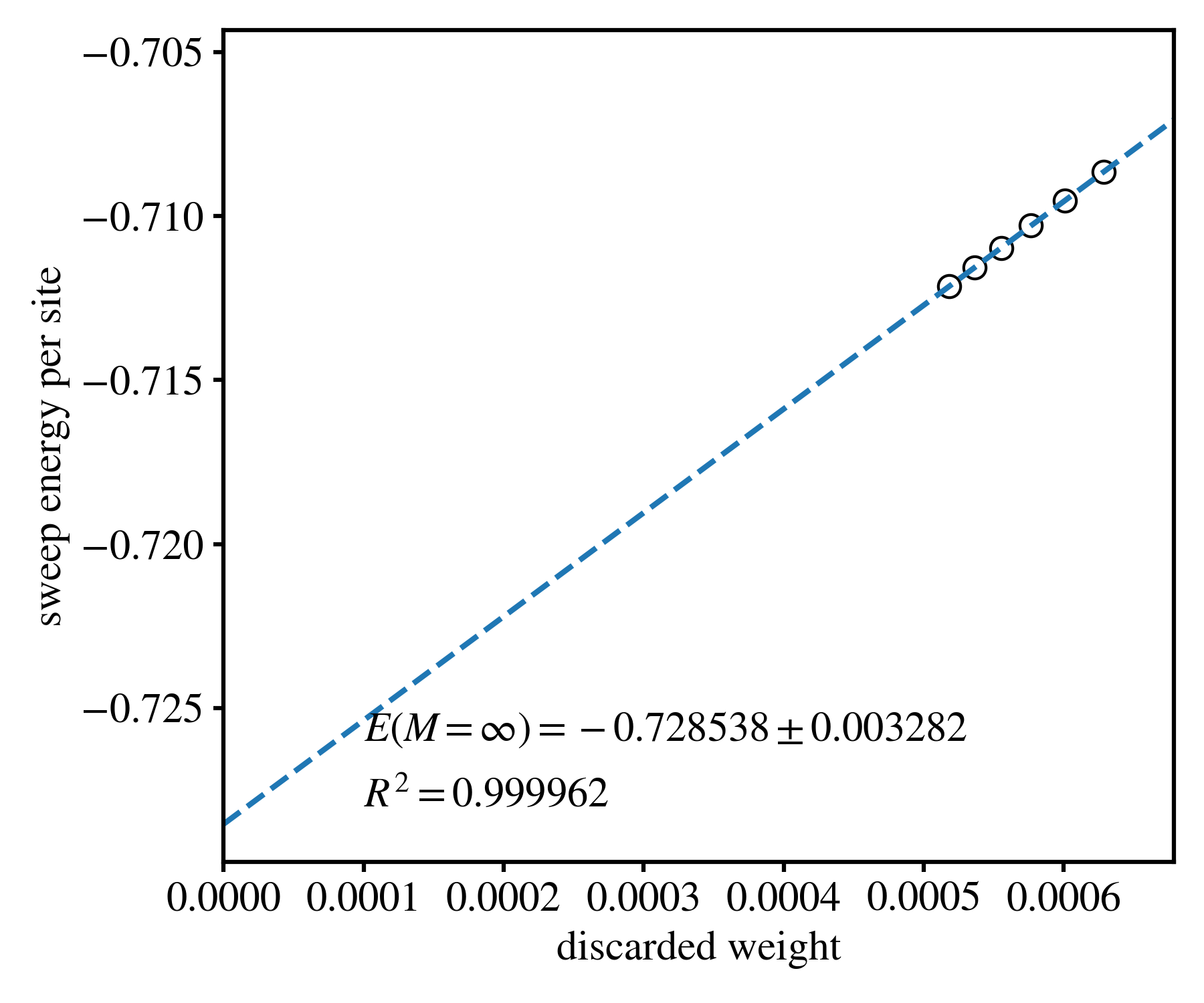}
    \caption{Extrapolation of the DMRG energy with respect to discarded weight. The circles correspond to the individual DMRG results.}
    \label{fig:DMRGextrap}
\end{figure}

The DMRG benchmark presented in Sec.~\ref{sec:2d_hub} was obtained by extrapolating the energy towards zero truncation error (see Fig.~\ref{fig:DMRGextrap}).  The accuracy of a finite bond dimension two-site DMRG calculation is measured by its truncation error which is a function of the number $m$ of retained density-matrix eigenstates  (for more details, see Ref.~\citenum{White93}). 
The error in the ground-state energy is proportional to the so-called discarded weight $W_m$,  defined as the total weight of the discarded density-matrix eigenstates:
\begin{equation}
W_m = \sum_{i = m+1}^d w_i.
\end{equation}
Here, $d$ is the dimension of the density matrix and $w_i$ is its $i$th eigenvalue. One can estimate the exact result by linearly extrapolating the DMRG energy to zero truncation error
$W_m \to 0$, for more details see Ref.~\citenum{legeza1996accuracy,chan2002highly,white2007neel}. In the computations presented here, the maximal bond dimension was 8000 (SU2 multiplets) corresponding to a discarded weight of less than $6\times 10^{-4}$. 

\section{Initial state dependence H$_{36}$}
\label{App:h36StateIndep}

\begin{figure}
    \centering
    \includegraphics[width= 0.55\textwidth]{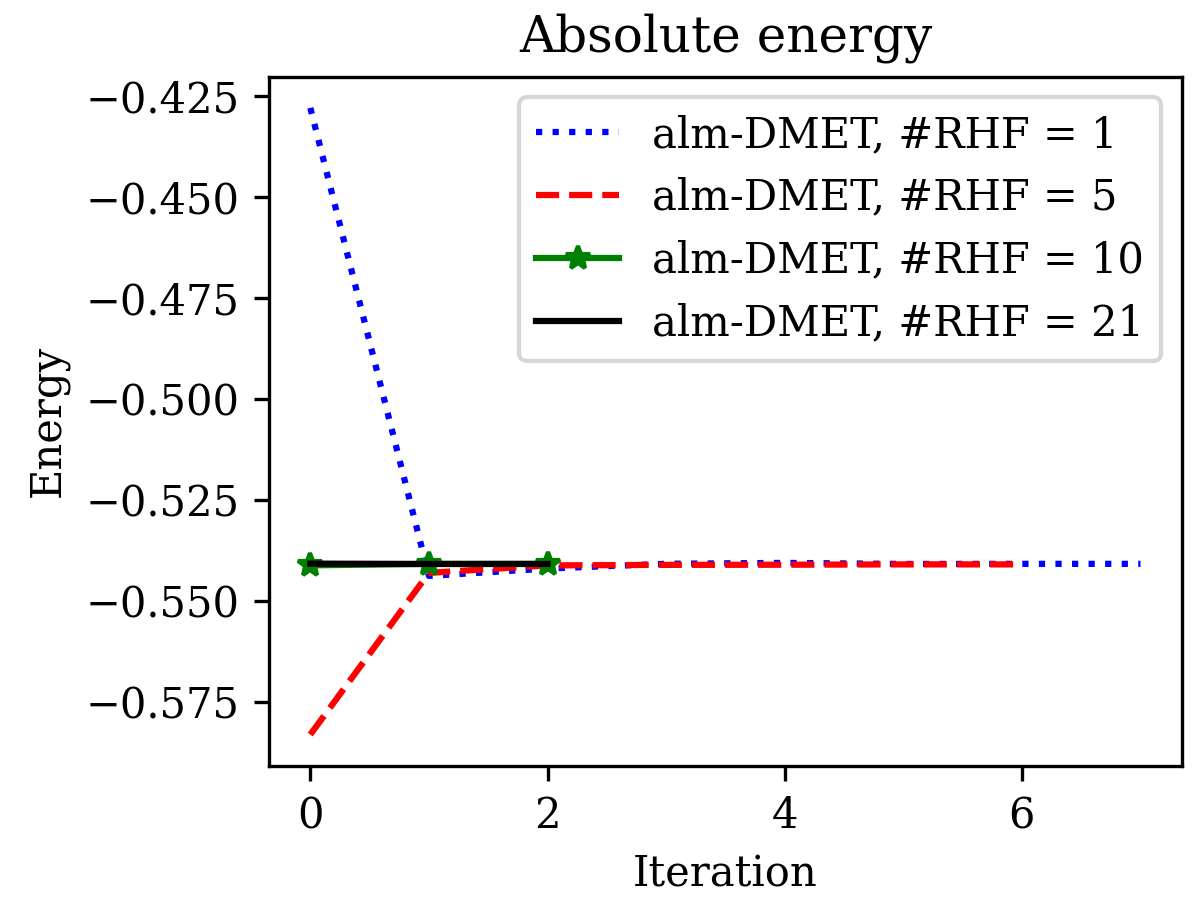}
    \caption{Convergence of the absolute energy of the linear H$_{36}$ system using alm-DMET with different initial guesses (RHF converged within 21 iterations).}
    \label{fig:H36_init_dep}
\end{figure}

For the H$_{36}$ example, we also investigated the initial state dependence of alm-DMET. We set the initial RDM to be the unconverged RHF solution after $1,5,10,21$ iterations, respectively and find that the solution of alm-DMET is independent of the initial guess, see Fig.~\ref{fig:H36_init_dep}.

\bibliography{ldmet.bib}

\end{document}